\input harvmac
\input epsf
\def\us#1{\underline{#1}}
\def\nup#1({Nucl.\ Phys.\ $\bf {B#1}$\ (}
\noblackbox
\Title{\vbox{
\hbox{HUTP-97/A090}
\hbox{\tt hep-th/9711013}
}}{Branes and Toric Geometry}

\bigskip
\centerline{Naichung Conan Leung$^{a}$ and Cumrun Vafa$^{b}$}
\bigskip
\bigskip\centerline{\it $^{a}$ School of Mathematics,
University of Minnesota, Minneapolis, MN 55455}
\centerline{\it $^{b}$Lyman Laboratory of Physics, Harvard
University,
Cambridge, MA 02138}

\vskip .3in
We show that toric geometry can be used rather effectively
to translate a brane configuration to geometry.  Roughly
speaking the skeletons of toric space are identified with
the brane configurations.  The cases where the local geometry
involves hypersurfaces in toric varieties (such as ${\bf P}^2$
blown up at more than 3 points)
presents a challenge for the brane picture.  We also find
a simple physical explanation of Batyrev's construction
of mirror pairs of Calabi-Yau manifolds using T-duality.

\goodbreak

\Date{\vbox{\hbox{\sl {November 1997}}}}

%
\parskip=4pt plus 15pt minus 1pt
\baselineskip=15pt plus 2pt minus 1pt
\newsec{Introduction}
One of the main new physical insights we have recently gained
from string theory is that geometric singularities of the internal
compactification manifold encode a great deal of information
about quantum
field theories.  Turning things around we can engineer quantum
field theories by suitably choosing singularities under consideration
and use them to gain insight into quantum field theories
(see for example
\ref\geoeng{S. Kachru and C. Vafa, \nup 450 (1995) 69\semi
S. Kachru et. al., \nup 459 (1996) 537\semi
A. Klemm, W. Lerche, P. Mayr, C. Vafa and N.
Warner,
\nup 477 (1996) 746 \semi
A. Brandhuber and S. Stieberger, Nucl. Phys. {\bf B488}
(1997) 199\semi
J. Schulze and N. P. Warner, hep-th/9702012 \semi
J. M. Rabin, hep-th/9703145.
 }\ref\kkv{S. Katz, A. Klemm and C. Vafa,
hep-th/9609239}\ref\kvv{S. Katz and C. Vafa, hep-th/9611090.}\ref\kmv{
S. Katz, P. Mayr and C. Vafa,
hep-th/9706110.}\ref\vom{C. Vafa, hep-th/9707131.}\ref\lerc{W. Lerche,
hep-th/9709146.}).  This program of
studying QFT's
is called {\it geometric engineering}.

On the other hand there has been another direction of
construction of field theories involving branes
(see for example \ref\bran{A. Hanany and E. Witten,
 hep-th/9611230}\ref\brref{E. Witten,
hep-th/9703166}).  Some of these cases are already
known to be equivalent, by T-dualities to the geometrical
cases (see for example \ref\klemet{A. Klemm, W. Lerche, P. Mayr, C. Vafa and N.
Warner,
\nup 477 (1996) 746.}\ref\oogv{H. Ooguri and C. Vafa, \nup 463
 (1996) 55}\ref\ovr{H. Ooguri and C. Vafa,
hep-th/9702180}\ref\hov{K. Hori, H. Ooguri and C. Vafa,
hep-th/9705220}).
Here we try to extend this dictionary to a more general class
of theories and in particular to 5 dimensional critical
theories constructed by Hanany and Aharony
\ref\hanah{O. Aharony and A. Hanany, hep-th/9704170}\
and studied further in \ref\kolet{B. Kol,
hep-th/9705031}\ref\yan{A. Brandhuber, N. Itzhaki, J. Sonnenschein, S.
Theisen and S. Yankielowicz, hep-th/9709010.}\ref\hanar{O. Aharony, A. Hanany
and B. Kol,
hep-th/9710116.}.
The approach
we follow will also lead to a simple geometric realization
of the Riemann surface for $N=2$ theories appearing
in \klemet\kmv\
through fivebranes of type IIA.
This is also related to the recent observations
made in \kolet \yan
\hanar \foot{
We note that the notion of `grid diagrams' discussed by
those authors in this context has been well known as a standard
construction in toric geometry and was extensively discussed
and used already in
\kmv .  In this paper we will also relate the physics of the situation
discussed in
\kmv\ to the brane realization of the same theories.}.
Even though here we limit ourselves to few examples, the approach
we take is quite general and can be applied to many other cases.

The summary of our results is as follows:  Toric geometry
involves viewing manifolds as roughly speaking
products of some space with a torus.  The only non-triviality
involves the fact that on some loci certain cycles of tori
can shrink.  Toric geometry is a way to encode this combinatoric
data as to which cycles shrink where.  This constitutes
faces of the polytope describing the toric spaces.
 On the other hand
vanishing cycles have been known to be associated with branes.
This connection thus identifies these toric skeletons directly
with branes of appropriate types!

Toric geometry, however, can be used in a more general
way to get interesting geometries, namely by going
to a higher dimensional space and imposing equations.
This presents a major challenge for the brane picture and it
is not clear how to modify the brane story to accommodate this simple
geometric idea.
This is precisely the flexibility that the geometric constructions enjoy
over the brane picture;  It would be interesting to try to find
a way of adapting the brane picture to such cases as well.

The organization of this paper is as follows:  In section two
we give a very simple overview of toric geometry (intended
for physicists unfamiliar with it).  In section 3 we describe
the relation between toric geometry and branes of various types.
Finally
in section 4 we use $R\rightarrow 1/R$ duality in the context of toric
geometry to give a simple intuitive explanation of Batyrev's construction
of Calabi-Yau mirror pairs \ref\baty{V. Batyrev, J.
Alg. Geom. $\us{3}$ (1994) 493.} (see also
\ref\roa{S.-S. Roan, Internat. J. Math. 2 (1991) 439.}).

\newsec{Review of Certain Aspects of Toric Geometry}
In this section we review certain aspects of toric geometry,
intended mainly for physicists unfamiliar with the ideas
in toric geometry.  We aim to give a very simple treatment
of the ideas of toric geometry.  For a detailed pedagogical
review emphasizing other aspects of toric constructions
see \ref\fult{W. Fulton, {\it Introduction to Toric Varieties},
Annals of Math. Studies, No. 131, Princeton University
Press, 1993.}\ref\mople{D.R. Morrison and M.R. Plesser, Nucl. Phys.
{\bf B440} (1995) 279.}.

Toroidal compactifications are among the most special
classes of compactifications in string theory. They
preserve the maximal amount of symmetry a lower dimensional
theory can possibly have starting from a higher dimensional one.
For example a d-dimensional torus admits a $U(1)^d$ translational
symmetry.
Even though it would be easiest to analyze the physical systems
under such compactifications, they would typically have
 too much
symmetry for many applications of interest in physics. The next best
thing in physics is compactifications
of something that comes close to being toroidal. Toric
geometry basically studies geometries where there is a $U(1)^d$ action,
as in the $T^d$ case, but unlike the toroidal case,
the $U(1)^d$ action is allowed to have fixed points.
The basic idea in characterizing such geometries is to isolate
the fixed point structure and use that to encode the geometry.

It is best to start with some examples:

Example i)  Consider the complex plane ${\bf C}$.  This manifold
admits a $U(1)$ action
$$z\rightarrow z {\rm exp}(i\theta )$$
with a fixed point at $z=0$.  The geometry of the plane
can be represented by a half-line, corresponding to $|z|$
above which there is a circle.  Moreover the circle shrinks
at the end of the half-line.


\bigskip
\epsfxsize 2.truein
\epsfysize 2.truein
\centerline{\epsfxsize 2.truein \epsfysize 2.truein\epsfbox{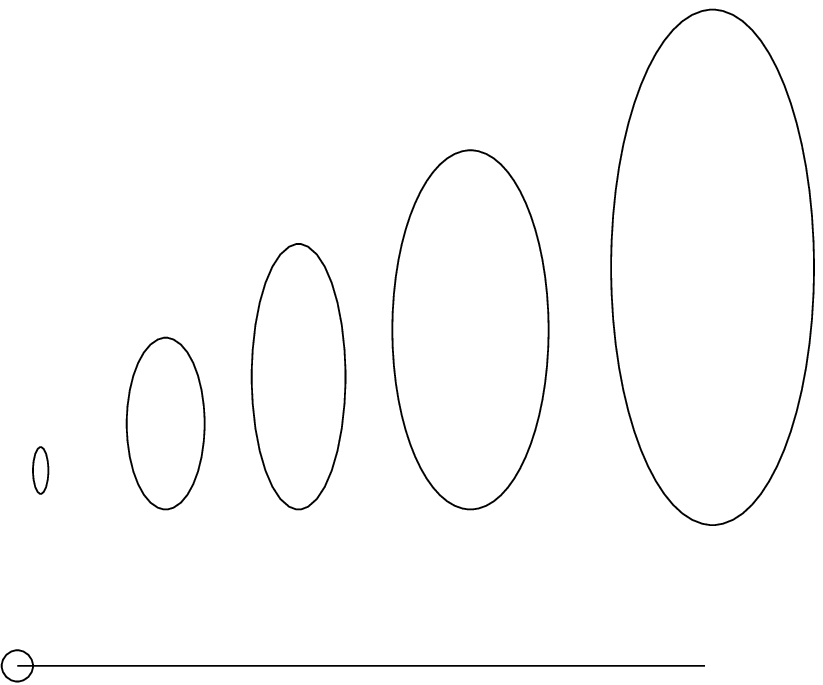}}
\leftskip 2pc
\rightskip 2pc
\noindent{\ninepoint\sl \baselineskip=8pt {\bf Fig.1}: {\rm
Complex plane can be viewed as a half-line with a circle on top,
which shrinks at the end.}}


Example ii) Consider the 2-sphere ${\bf P}^1$, which can be
viewed as the compactified complex plane $z$.  Again there is
a $U(1)$ action, just as above, in terms of which the 2-sphere
can be represented as an interval times a circle, where the
circle shrinks at the two ends, corresponding to north and south poles
of the sphere.  The coordinates on the interval can be identified
with a function of $r=|z|$.
The length of this interval is determined by the size of the 2-sphere.
More precisely, if
the 2-sphere has a metric which is $\alpha $ times the standard metric on
the 2-sphere, namely the Fubini-Study metric ${{\left| dz\right| ^2}\over {
( 1+\left| z\right| ^2) ^2}}$ then the coordinate on the interval
is given by
$$x={{\alpha \left| z\right| ^2}\over {( 1+\left| z\right|
^2)}}$$
 which runs from $0$ to $\alpha .$ Notice that the origin
of the interval is not relevant and we can perform a coordinate transformation
of the
interval by translation $x\rightarrow x+x_0$ without changing our picture.


\bigskip

\centerline{\epsfxsize 2.truein \epsfysize 2.truein\epsfbox{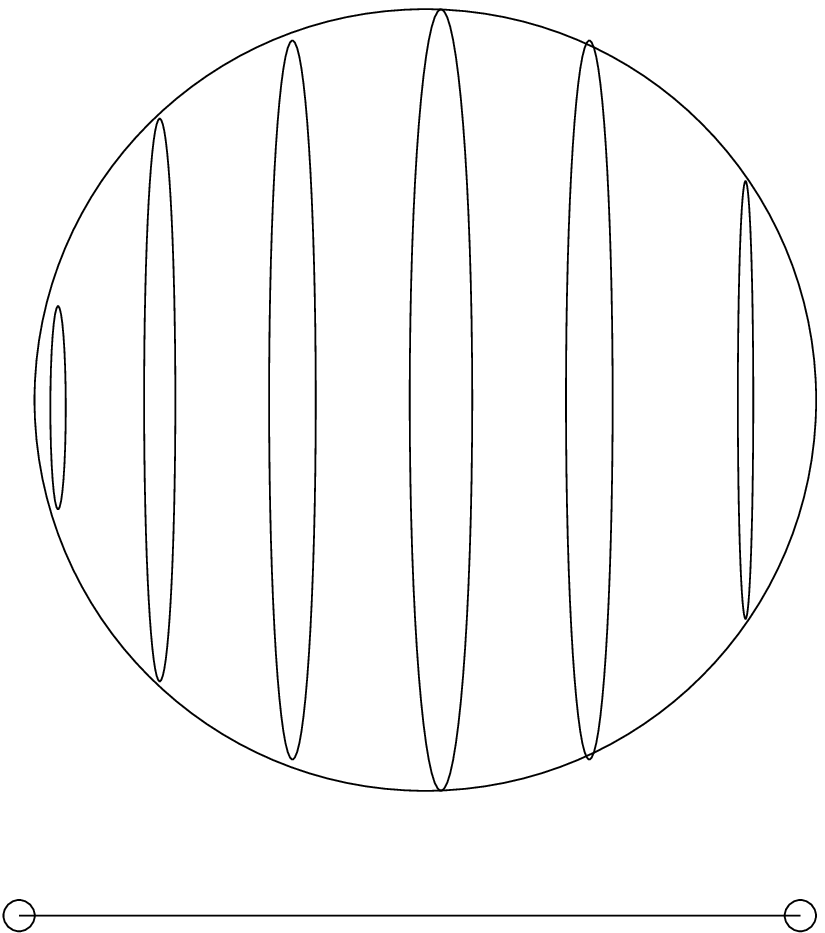}}
\leftskip 2pc
\rightskip 2pc
\noindent{\ninepoint\sl \baselineskip=8pt {\bf Fig.2}: {\rm
The 2-sphere can be viewed as an interval with a circle
on top, where the circle shrinks to zero size at the two ends.}}


In various applications it
is important to consider in addition a holomorphic line
bundles on the 2-sphere.
  Then the first Chern
class $c_1$ of the bundle is a $(1,1)$ form which can be taken
to be the volume element corresponding to the Kahler form for the
Fubini-Study metric $c_1=n$.  In this case, however, we will have to use
{\it integral} $\alpha$ giving integral volume of the sphere because
$$\int_{{\bf P}^1}c_1 =n$$
for some integer $n$ and we can identify $\alpha =n$.  If we wish
to consider the sphere together with the bundle on top it is then convenient
to choose the interval to go from $0$ to $n$.
Such an interval will be
called integral.
Note that the number of holomorphic sections of the bundle is then related to
the number of integral points on the interval.  Thinking
of the volume form on the 2-sphere as a symplectic form, with $x$ giving
the radial direction, the number of sections of the bundle is related
to the dimension of the Hilbert-space, which is the same as the
 Bohr-Sommerfeld quantization rule for a compact phase space.
Note that one can identify the holomorphic sections in this
case with $z^i$ where $i=0,...,n$.


\bigskip

\centerline{\epsfxsize 2.truein \epsfysize .5truein\epsfbox{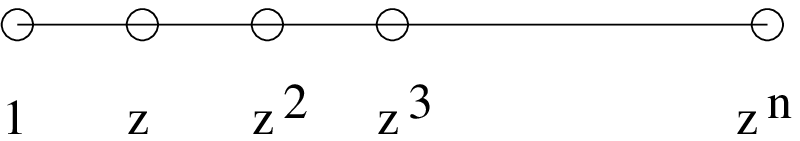}}
\leftskip 2pc
\rightskip 2pc
\noindent{\ninepoint\sl \baselineskip=8pt {\bf Fig.3}: {\rm The integral
lattice can be used to summarize the information about sections
of bundle on the sphere.  Each point on the lattice corresponds
to a section of the bundle.}}


%

Note that near each of the two ends the geometry is the same
as the example 1, which is just the statement that near the north pole
or the south pole we can view the sphere as a patch which is just
${\bf C}$.

Example iii) Our next example is ${\bf P}^2$.  As is well known
${\bf P}^2$ is the space of three complex numbers $(z_1,z_2,z_3)$
not all zero, modulo identifying them
up to multiplication by a non-zero complex
number.  In this case we have a $U(1)^2$ action, consisting
of the $U(1)^3$ action on the phases of the $z_i$, modulo
the action of the diagonal $U(1)$ which acts trivially
on ${\bf P}^2$.  We can consider a basis of the $U(1)^2$ action
to consist of
$$ (z_1,z_2,z_3)\rightarrow (z_1 {\rm exp}(i \theta),z_2
{\rm exp}(i \phi),z_3)$$
The fixed point of $\theta $ action consists of $(0,z_2,z_3)$ up
to an overall rescaling, which gives a ${\bf P}^1$ parameterized
by $z_2/z_3$.
Similarly the fixed point of the $\phi$ action is a ${\bf P}^1$
parameterized by $z_1/z_3$.  Also if we consider $\theta =\phi$
diagonal $U(1)$ we get another fixed point locus being the ${\bf
P}^1$ parameterized by $z_1/z_2$.  Moreover, there are
three fixed points where both $U(1)$ actions have fixed points
(when any pair of $z_i,z_j=0$) corresponding to the intersection
of any pair of these ${\bf P}^1$'s.    We are interested
in viewing ${\bf P}^2$ as a space having generic $T^2$ fibers
parametrized by the action of $(\theta,\phi )$ introduced above.
We have to choose coordinates for a two real dimensional
base which is invariant under the $T^2$ action.  Note that
$r_1=|z_1/z_3|$ and $r_2=|z_2/z_3|$ are such coordinates.
Again we can map it to finite regions by considering
appropriate functions of $r_1,r_2$.
We thus can represent the ${\bf P}^2$ according to the figure:
%

\bigskip

\centerline{\epsfxsize 2.truein \epsfysize 2.truein\epsfbox{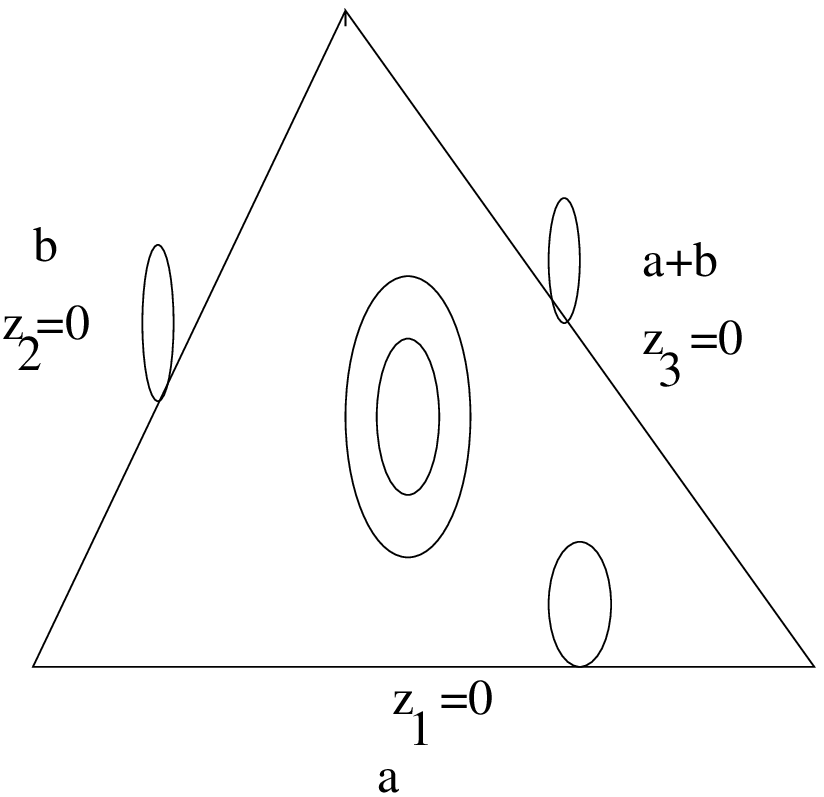}}
\leftskip 2pc
\rightskip 2pc
\noindent{\ninepoint\sl \baselineskip=8pt {\bf Fig.4}: {\rm The toric
realization of ${\bf P}^2$ involves a triangle over each
point of which there is a 2-torus, which shrinks to a circle
at each edge, and where it shrinks to a point at each vertex of the
triangle.  Each edge of the triangle, with the circle on top, corresponds
to a ${\bf P}^1\in {\bf P}^2$.}}

%
where this represents the base of the ${\bf P}^2$. Above
each point in the interior of the triangle we have a
$T^2$ fiber $(\theta,\phi )$.  Let us denote the cycles
of this torus corresponding to $\theta, \phi$ by $a,b$
respectively.  The $T^2$ fibration degenerates
near the edges of the triangle, where over one edge $a$
shrinks (corresponding to $z_1=0$), over the other $b$ shrinks
(corresponding to $z_2=0$) and over the
other $a+b$ shrinks (corresponding to $z_3=0$).  On the vertices
of the triangle both $a$ and $b$ shrink.
For various applications
it turns out to be convenient to introduce the following construction.
One realizes the base of the ${\bf P}^2$ in ${\bf R}^2$, where
we orient each face so that the normal vector to that face corresponds
to the cycle direction of the fiber $T^n$ which it shrinks.  For example
for the ${\bf P}^2$ example above we draw the ${\bf P}^2$ base as follows:


\bigskip

\centerline{\epsfxsize 2.truein \epsfysize 2.truein\epsfbox{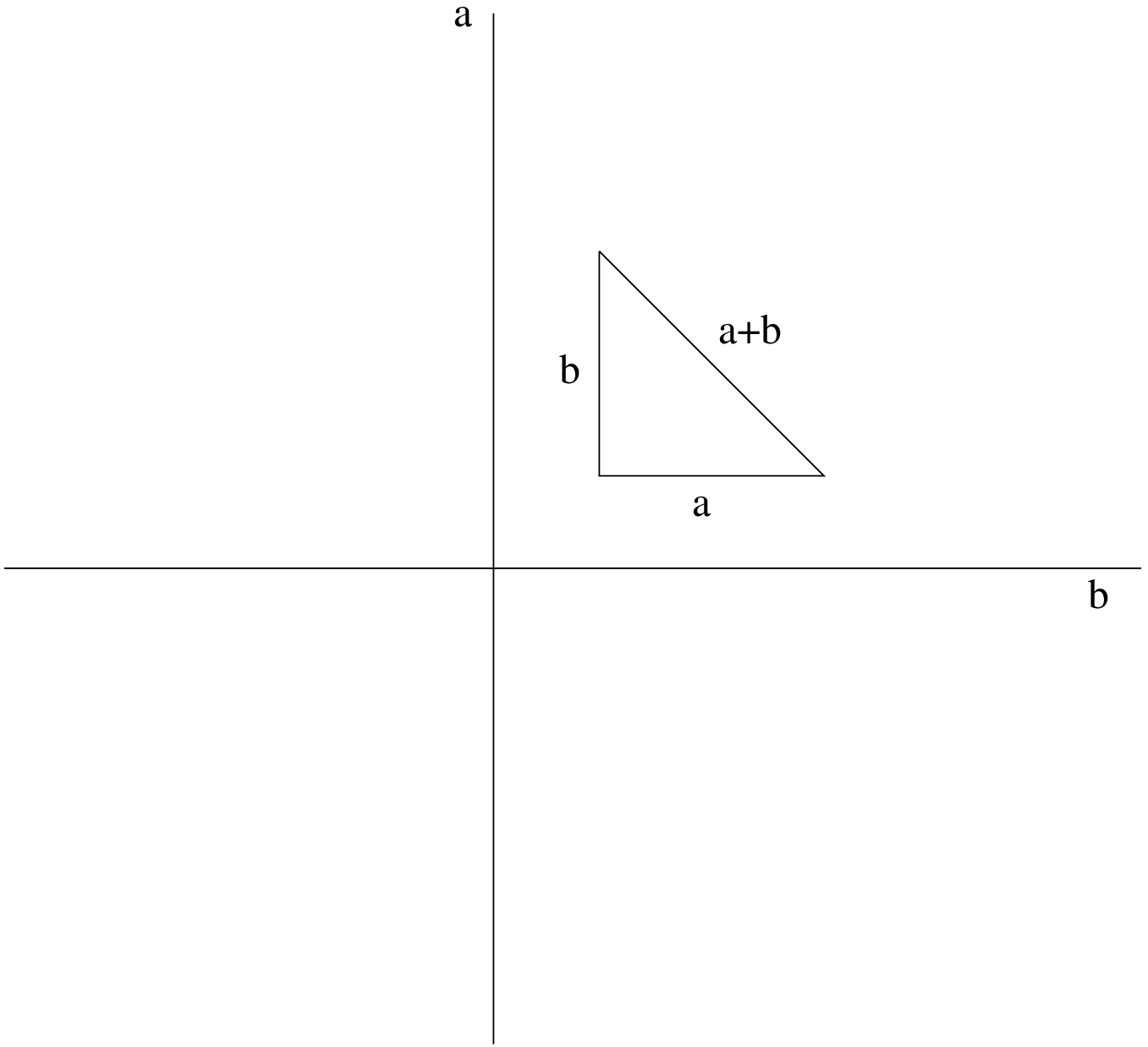}}
\leftskip 2pc
\rightskip 2pc
\noindent{\ninepoint\sl \baselineskip=8pt {\bf Fig.5}: {\rm It
is natural to draw the faces of the toric space angled in such
a way that is normal to the direction of the cycle which vanishes
over it.}}


Now if we wish to emphasize the bundle structure, all we have
to do is to choose the vertices to lie on integral points ${\bf Z}^2$.  Given
the geometry of ${\bf P}^2$, the angles of all the three edges are
fixed and all we can vary is an overall size.  This is in accord with
the fact that line bundles on ${\bf P}^2$ are characterized by
the choice of an integer $n$.
 In this case the line bundle restricted
to each of the three ${\bf P}^1$'s will correspond to the number of
integral points on the interval and is identified with this $n$.  Moreover, the
totality of the
points in the triangle (the integral points in the interior
as well as points on the boundary) will correspond to the number of holomorphic
sections of the bundle over ${\bf P}^2$.  In fact if we denote
a point in the triangle by $(a,b)$ we associate to this the
section $z_1^az_2^b$ (note that $a+b\leq n$, $a\geq 0$ and $b\geq 0$).
The example of ${\bf P}^2$ with degree $3$ bundle is shown below:
%

\bigskip

\centerline{\epsfxsize 2.truein \epsfysize 2.truein\epsfbox{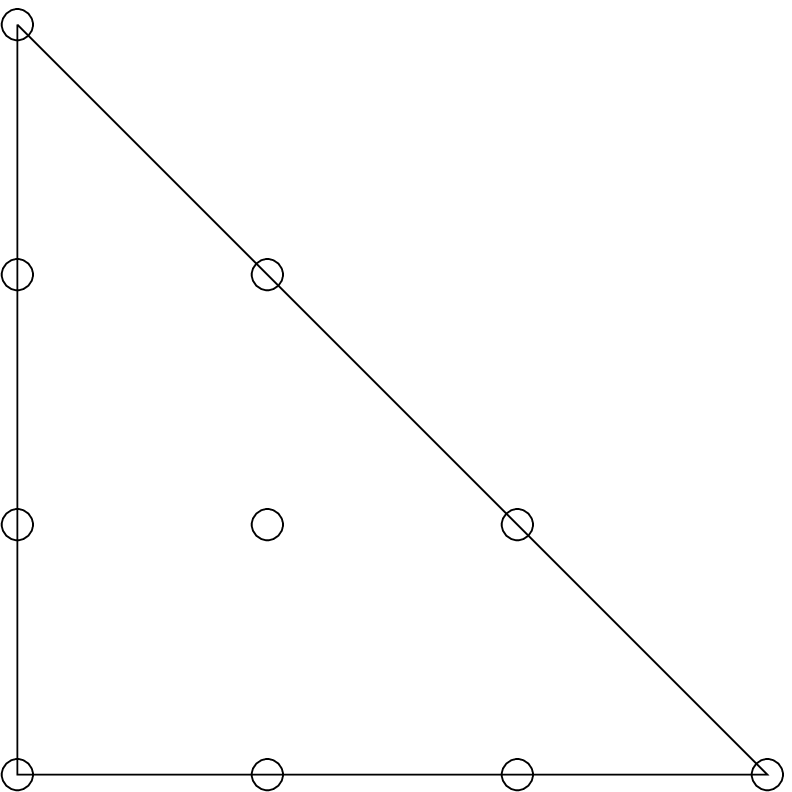}}
\leftskip 2pc
\rightskip 2pc
\noindent{\ninepoint\sl \baselineskip=8pt {\bf Fig.6}: {\rm The
integral toric realization of ${\bf P}^2$ with a line bundle
of degree $3$.  All the
integral points shown in the figure correspond to sections of that bundle.
}}


%

Example iv)  We can also describe ``blowing up" of
${\bf P}^2$ at some number of generic points $n\leq 3$ in a similar
manner.
What blowing up means in this context is to replace a point
on ${\bf P}^2$ by a sphere ${\bf P}^1$.  With no loss of generality
we can take the point to be at any of the three vertices of the
toric triangle (by the $SL(3)$ symmetry of ${\bf P}^2$).
Since blowing up means replacing a point by a ${\bf P}^1$ and
that is realized in toric language by an interval, as discussed
in example ii), this implies that ${\bf P}^2$ with one point
blown up will be torically given by:
%


\bigskip

\centerline{\epsfxsize 2.truein \epsfysize 2.truein\epsfbox{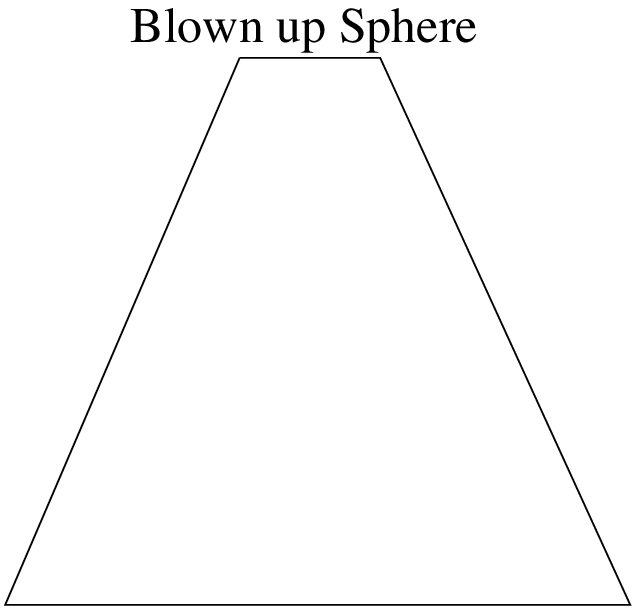}}
\leftskip 2pc
\rightskip 2pc
\noindent{\ninepoint\sl \baselineskip=8pt {\bf Fig.7}: {\rm
Blowing up can be realized very easily using toric geometry.
Here we are drawing the blowing up of ${\bf P}^2$} at one point
(what used to be the top vertex of ${\bf P}^2$) which has
the effect of replacing it by a ${\bf P}^1$ (shown as the top
interval in the above figure.  The size of the interval
is a direct measure of the size of the blown up ${\bf P}^1$.}

%
where again one can work out what cycle of $T^2$ vanishes over the new
${\bf P}^1$.  Note that blowing up, up to three generic points
can be realized torically, because using $SL(3)$ symmetry of ${\bf P}^2$
we can map any three points to the three vertices of the triangle
above.  Beyond three points we can still blow up, but that
cannot be realized torically for generic points.  Only if we choose
special points which lie at the corner of the toric base can
we continue blowing up torically.
However it is known that the manifold
one gets by blowing up more than 3 points on ${\bf P}^2$  will
depend on where the points are chosen.  If they
are done generically, then we get what is called a del-Pezzo
surface (up to blowing up 8 points) and has a positive
first Chern class $c_1>0$.  But if the points are not generic,
as will be the case in the toric realization where we choose
more than 3 points to blow up,
the manifold we get will not have $c_1>0$.  This will prove
important for certain comparisons with physical realization
via 5-branes which we will discuss later in this paper.
We emphasize that this is a limitation
of toric realization of blown up ${\bf P}^2$ and not a
reflection of any intrinsic problem with the geometry of
blown up ${\bf P}^2$.

Example v)  We can now generalize easily  to ${\bf P}^n$,
where we have an $n$-dimensional toroidal fiber $T^n$
over an $n$-dimensional base, which is identified with an
$n$-dimensional simplex. Let us denote
a basis of the cycles of $T^n$ by $a_1,...,a_n$.  The
simplex has $n+1$ boundary faces, over each of which a 1-cycle
of $T^n$ shrinks.  These can be taken to correspond to
$a_1,...,a_n,\sum a_i$.


\bigskip

\centerline{\epsfxsize 2.truein \epsfysize 2.truein\epsfbox{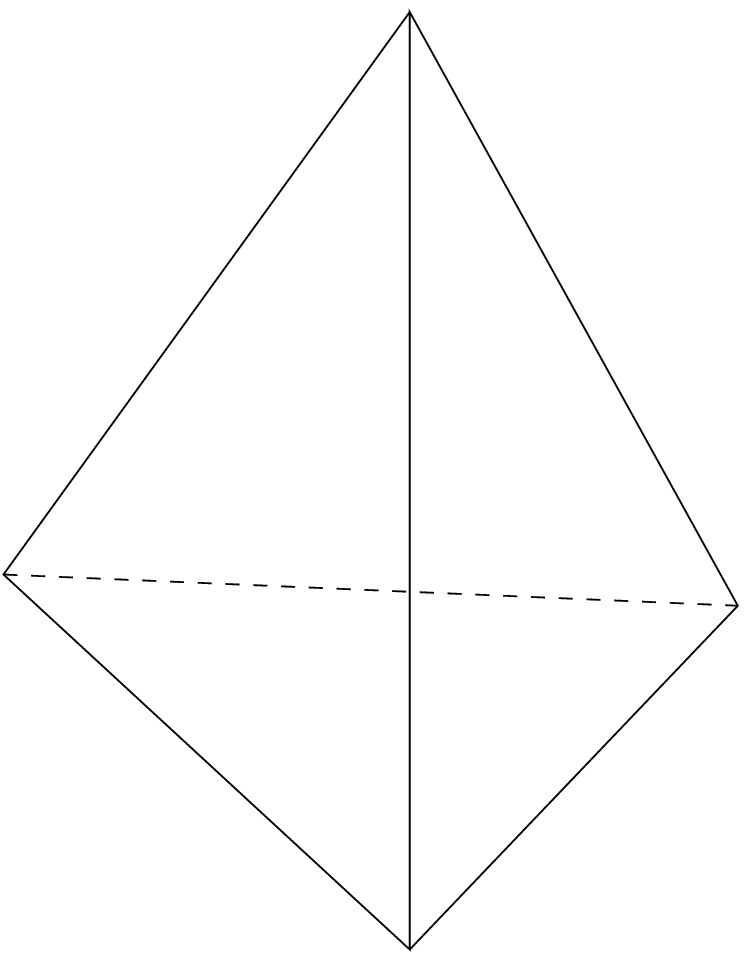}}
\leftskip 2pc
\rightskip 2pc
\noindent{\ninepoint\sl \baselineskip=8pt {\bf Fig.8}: {\rm Here
we are showing the toric realization of ${\bf P}^3$.}}


Each face is an $n-1$ dimensional simplex.  Two such faces
meet over an $n-2$ dimensional simplex, over which two
cycles of the $T^n$ shrink.  More generally $k$ such
faces meet over an $n-k$ dimensional simplex over which
$k$ cycles of $T^n$ shrink.  In particular when $n$ of
them meet (which happens at $n+1$ points) the whole
$T^n$ fiber has shrunk.

Example ii') The examples of ${\bf P}^n$ discussed above cannot
be used for string compactification as they are not solutions
to Einstein's equations (they are not Ricci-flat).  However they
can be part of a local geometry of a Calabi-Yau near a singularity.
For example consider the case of ${\bf P}^1$.  It is known
that the cotangent space $T^*{\bf P}^1$ can appear as part of
Calabi-Yau compactifications (for example near an $A_1$
singularity of $K3$).  This space is also toric.  If we
denote the coordinates of $T^*{\bf P}^1$ by $(z,p)$
corresponding to ${\bf P}^1$ and the cotangent direction
respectively, we can consider two circle actions
on this space.  The first one is the one induced
from the action on the ${\bf P}^1$ base to the normal
direction (taking into account
that $pdz$ is invariant)
$$(z,p)\rightarrow ({\rm exp}(i\theta )z,{\rm exp}(-i\theta )p)$$
and the other circle action is new and acts entirely on the fiber
$$(z,p)\rightarrow (z, {\rm exp}( i\phi)p)$$
Let the $(a,b)$ cycles of the $T^2$ to correspond
to the $(\theta, \phi )$ actions respectively.  Then as before
we can use $|z|$ and $|p|$ as defining a base for a $T^2$
fibration, with a $T^2$ fiber corresponding
to the $(\theta, \phi)$ action.
There are three fixed loci of this toric action over which
an $S^1$ in the fiber shrinks.  These
 correspond
to $z=0,z=\infty $ and $p=0$ (note that the $p$ direction
is non-compact and so it does not have a point at infinity).
  At $z=0$ the invariant direction corresponds
to setting $\theta =\phi$.  In other words the
$a-b$-cycle shrinks.  At $p=0$ the $b$-cycle shrinks.
To find out what the toric action is near $z=\infty$
it is convenient to change the patch to $\tilde z=1/z$.
Noting that $pdz$ is invariant this undergoes
a transformation $\tilde p=-pz^2$, in terms of which the
$T^2$ action becomes
$$({\tilde z},{\tilde p})\rightarrow ({\rm exp}(-i\theta )
{\tilde z},{\rm exp}(i\phi +i\theta){\tilde p})$$
Thus at ${\tilde z}=0$, if we set $\phi=-\theta$, the action
on $T^*{\bf P}^1$ is trivial.  In other words at ${\tilde z}=0$
the cycle $a+b$ shrinks.  Thus the base of the toric
fibration is given by the geometry below:
%


\bigskip

\centerline{\epsfxsize 2.truein \epsfysize 1.5truein\epsfbox{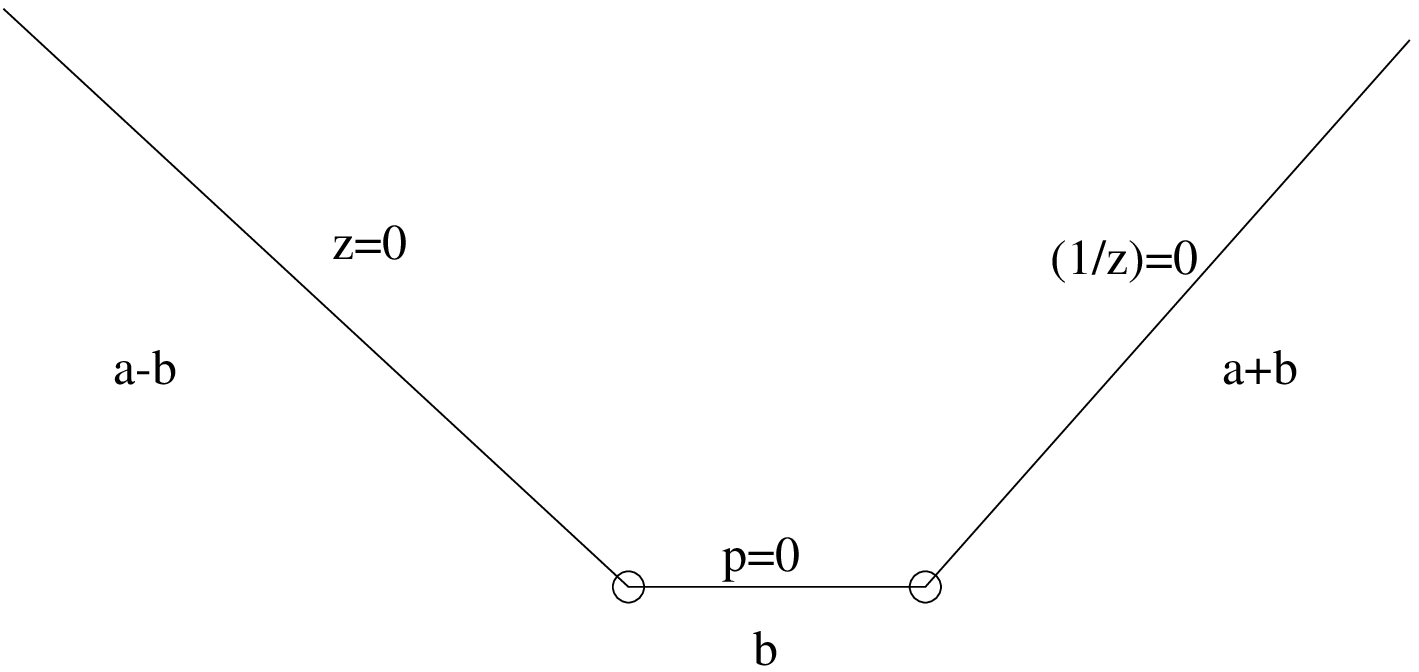}}
\leftskip 2pc
\rightskip 2pc
\noindent{\ninepoint\sl \baselineskip=8pt {\bf Fig.9}: {\rm The
toric realization of the blowing up of $A_1$ singularity
in $K3$.  The finite interval represents the blown up ${\bf P}^1$.
 Note that a half line emanating from the interval
going to the infinity corresponds to the cotangent bundle
at that point, which is a copy of the complex plane ${\bf C}$.}}



Note that the boundary of this figure corresponds
from left, to the $|p|$ over $z=0$, the $p=0$ which
corresponds to the $z$-sphere and is represented
by the interval, and finally the cotangent
direction over ${\tilde z}$.  The lines emanating
from the interval corresponding to the $z$-sphere
going to infinity correspond to the
non-compact cotangent direction over sphere.
It is also easy to generalize this to when we have
an $A_{n}$ singularity.  The geometry consists of
$n$ spheres intersecting according to the $A_n$ Dynkin
diagram.  The toric geometry is summarized as follows:
%


\bigskip

\centerline{\epsfxsize 2.truein \epsfysize 1.5truein\epsfbox{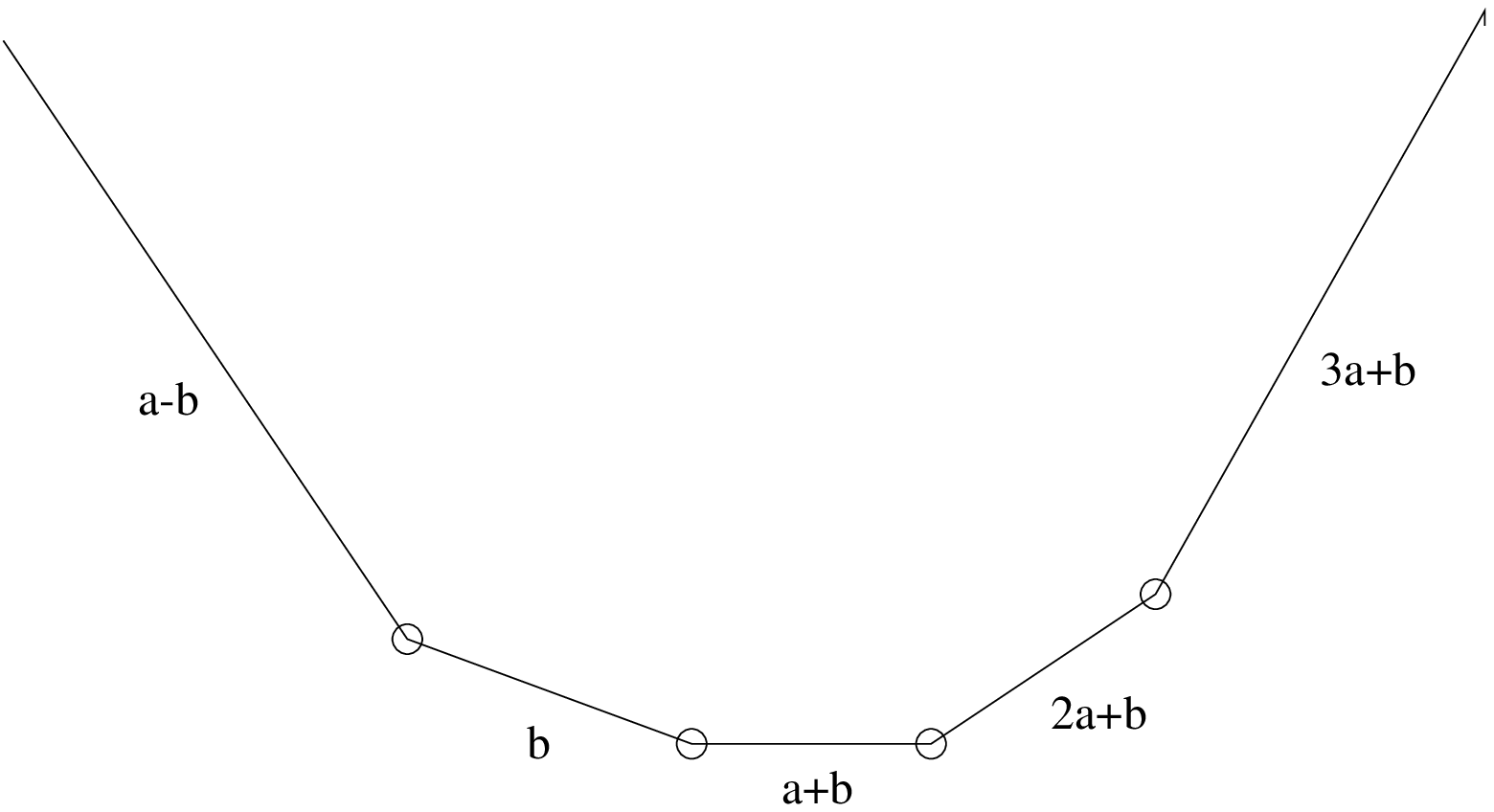}}
\leftskip 2pc
\rightskip 2pc
\noindent{\ninepoint\sl \baselineskip=8pt {\bf Fig.10}: {\rm
Shown in the figure is the blown up singularity $A_3$.  Note that
the three finite size intervals in the middle denote the three blown
up ${\bf P}^1$'s.  Also note that the Dynkin diagram of $A_3$ is
visibly seen here by the intersection of neighboring ${\bf P}^1$'s
at a point.}}


%
Note that each ${\bf P}^1$ is given by an interval
in the above figure, and the size of the ${\bf P}^1$ is
represented by the size of the interval. In the limit
that a ${\bf P}^1$ shrinks, the interval shrinks
and we obtain a singular geometry. The geometry we have
depicted above is a smooth geometry corresponding to
``blowing up" the $A_n$ singularity.

Example iii)' Similar to the above example, a ${\bf P}^2$ can appear in a
Calabi-Yau, where there are some extra dimensions.  In particular
if this is embedded in a Calabi-Yau threefold there is a normal direction
which corresponds to a line bundle on ${\bf P}^2$.  The condition
that $c_1=0$ for the threefold implies that the normal bundle
is the canonical line bundle (corresponding to $(2,0)$ forms
on ${\bf P}^2$), thus cancelling
$c_1$ for the ${\bf P}^2$.  We will thus now have a 3-dimensional
local toric geometry, where the extra circle action comes from
the rotation on the phase of the normal line bundle.  Going through
the exercise just as we did for the case of ${\bf P}^1$ will give
the toric data summarized below:


\bigskip

\centerline{\epsfxsize 2.truein \epsfysize 2.truein\epsfbox{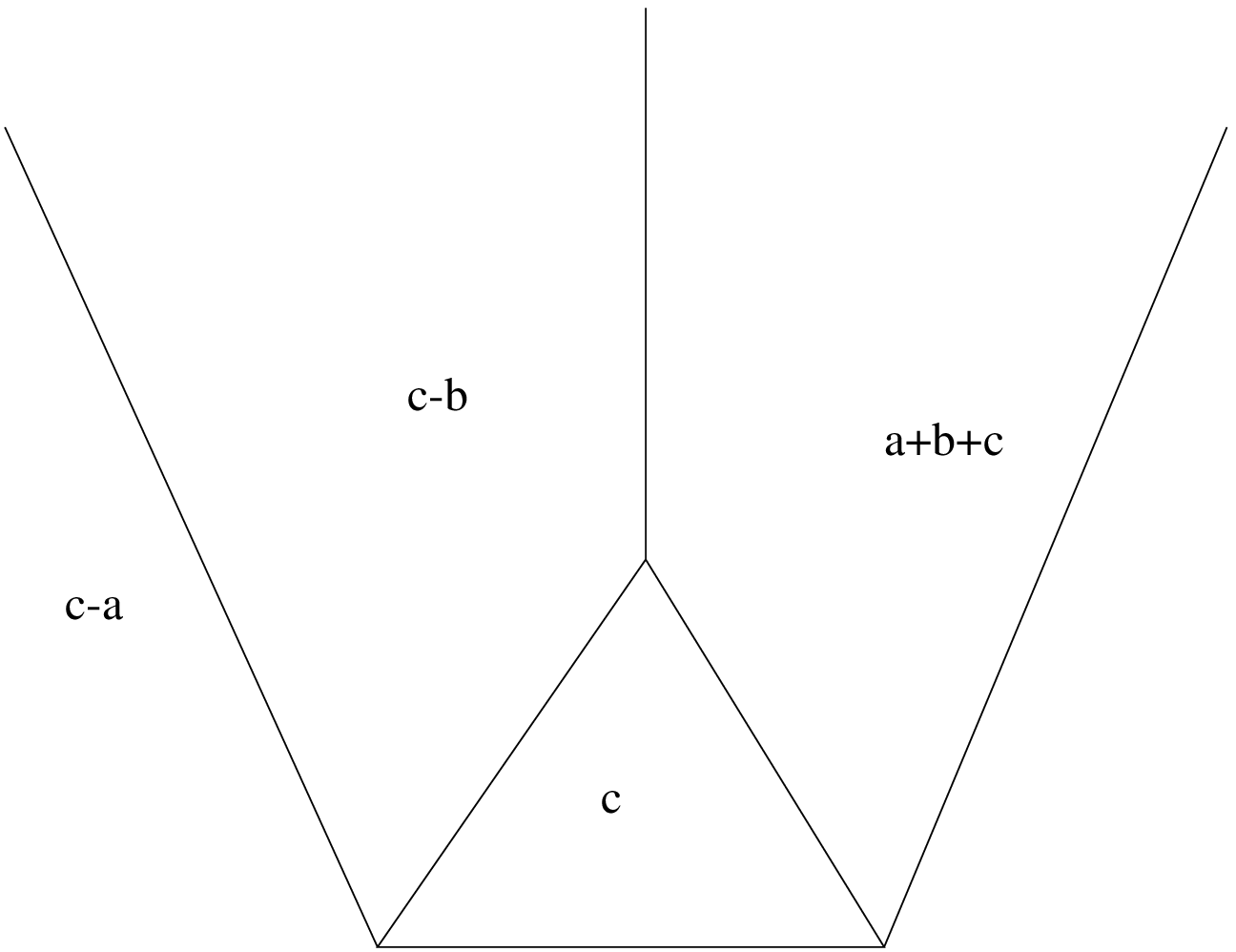}}
\leftskip 2pc
\rightskip 2pc
\noindent{\ninepoint\sl \baselineskip=8pt {\bf Fig.11}: {\rm The
toric realization of $N({\bf P}^2)$.  A copy of ${\bf P}^2$
is visible as the triangle at the bottom.  Each half line
emanating from any point on it, will correspond to the normal
direction of ${\bf P}^2$ in the Calabi-Yau threefold.}}


A copy of ${\bf P}^2$ is recognized at the bottom of the
above figure and the lines over it correspond to the normal
direction on ${\bf P}^2$.  If we call the extra circle direction
$c$, then the zero section of the normal bundle, which gives
a copy of ${\bf P}^2$ corresponds to $c$ being shrunk.  Similarly
the cycles that vanish at the other faces can also be worked out and
give the picture above.  This example can be easily generalized
to the case where we have a ${\bf P}^n$ sitting in an $n+1$
CY manifold, where again the normal direction to ${\bf P}^n$ is
identified with the space of $(n,0)$ forms on ${\bf P}^n$.

\subsec{Toric Varieties}
{}From the above examples it should be clear how to generalize
the notion of ${\bf P}^n$ or normal bundles to them, to a
more general class involving manifolds which admit toric action
\ref\toric{V. Guillemin, Moment Maps and Combinatorial Invariants of
Hamiltonian $T^n$-spaces, Birkhauser (1994)}.
We are interested in manifolds admitting a $T^n$ action, with
an $n$-dimensional base.  The $n$-dimensional base will have
$n-1$ dimensional boundary decomposed to various faces
where a particular $S^1$ cycle of the fiber shrinks, corresponding
to where the $T^n$ action has fixed loci.    Moreover when
$k$ of these faces meet a $T^{k}$ has shrunk to zero size.
The data defining the toric variety is precisely how these
faces meet and which cycles vanish over which face.

In general toric varieties will have singularities.  For example if you
consider the case of $A_{m-1}$ space, when we take the $m-1$ finite
size intervals
to zero size geometrically we get a singular space.  Torically the way
to read this singularity is rather simple:  The two edges
that now meet correspond to shrinking $b-a$ and $b+(m-1)a$ cycles.
Note that the lattice of 1-cycles on $T^2$ are {\it not} generated
by $b-a$ and $b+(m-1)a$ for $m\geq 1$.  Note that this is precisely
the case where we have a ${\bf C}^2/{\bf Z}_m$ singularity.
If we blow up the $m$-spheres
then it is easy to see that whenever two edges meet the vanishing cycles
form a basis for the lattice.  This turns out to be the general
consideration for a non-singular toric variety, namely
whenever $n$ faces meet we should get a basis for
the $n$ dimensional lattice dual to the $T^n$ fiber coming
from the vanishing cycles on each face.
To be more precise, if we denote the lattice generated by all one cycles in
the $T^n$ fiber by $M$ and we denote the sublattice generated by those
shrinking one cycles at any face of the polytope by $M_0$. Both $M$ and $M_0$
will have the same rank when the face is of zero dimensional, namely a
vertex. In this case the quotient $M/M_0=G$ will be a finite group of
reflexions.  Locally the geometry looks
like ${\bf C}^n/G$ which is singular at the origin
unless $M$ and $M_0$ are the same.
 In the case of $A_{m-1}$ singularity we have
$G={\bf Z}_m$.
 It is not difficult to
see that every points in a face which is adjacent to a smooth vertex point
is also a smooth points. Therefore to check smoothness of the toric
variety, it is sufficient to check only those vertex points.

\subsec{Hypersurfaces in Toric Varieties}

So far we have talked about the manifolds being
the toric varieties themselves.  However many interesting
geometries are not of this type.  For example, no
compact Ricci-flat manifold is toric--the above examples
gave some non-compact examples of Ricci-flat manifolds.
In order to remedy this, but still be close to the nice toric situation
one can start with a higher dimensional toric variety
and impose some equations.  The simplest set of such
cases involve degree $n+2$ hypersurfaces in ${\bf P}^{n+1}$ manifolds
which give rise to Calabi-Yau n-folds.  We have a polynomial
$$W(z_i)=0$$
where $i$ runs from 1 to $n+2$ and this is a homogeneous
equation of degree $n+2$.  Note that we can view $W$ as a section
of a line bundle on ${\bf P}^{n+1}$ (of degree $n+2$); as discussed
before it is natural to associate in such cases an integral
polytope which has in addition the information of the monomials
$W$ in it.  For every point on the integral
polytope and its interior we can write a monomial deformation
for $W$.  What this means is as follows:  Consider for each
point $r=(r_1,...,r_{n+1})$ in the integral polytope, a monomial
$z^r=z_1^{r_1}... z_{n+1}^{r_{n+1}}$.  Then the manifold hypersurface
is described by an equation
\eqn\patch{\sum_{r\in {\rm polytope} }a_r z^r =0}
for some coefficients $a_r$,
where this is a local description of the manifold in a patch.
In particular shifting the points by an integral shift does not
change the local geometry (as long as we keep away from
$z=0, \infty$).  It is often convenient to choose the integral
polytope to contain the origin, in which case there would exist
a monomial deformation corresponding to addition of $1$ to the equation.

 Note that this hypersurface given by $W=0$ will not
have any toric symmetry, because for a generic choice
of $W$ the torus actions is not compatible with the equation--
i.e. $W$ is not invariant under them.  There is, however, a degenerate
limit of $W$ in which the space does become toric.
Consider in the homogeneous variables
$$W(z_i)=W_0(z_i)+\psi z_1 z_2\cdot \cdot \cdot z_{n+2}=0.$$
The deformation corresponding to $\prod z_i$ in the above homogeneous
variables gets mapped to the deformation given by 1 in the above
patch description of the manifold \patch .
Now we consider the limit $\psi \rightarrow \infty$.  In this limit
the equation for the hypersurface becomes approximately
$$\prod_i z_i=0.$$
This consists of the union of the boundary faces of the polytope
each of which corresponds to $z_i=0$.  Thus roughly
speaking the Calabi-Yau $n$-fold consists of an $n$ dimensional
real base over which we have $T^n$ fibers (one circle
has already shrunk on each face $T^{n+1}\rightarrow T^n$).
Note that in this limit where $ k$ faces of the polytope meet
the fiber is $T^{n+1-k}$.

More generally we can obtain a Calabi-Yau space
as a hypersurface in toric variety by considering what is known
as ``reflexive polytope" as we will explain now. Instead of the standard
$n+1$ simplex which corresponds to ${\bf P}^{n+1}$, we can use arbitrary
 polytope $\Delta$ in ${\bf R}^{n+1}$ to construct a corresponding
toric variety ${\bf P}_{\Delta}$ which might be singular. Each of the boundary
faces
of $\Delta$ will give a hypersurface in  ${\bf P}_{\Delta}$ which itself is a
toric variety
of one lower dimension. Equivalently we can view it as the zero locus of a
section $s$
of a line bundle $L$ corresponding to this face. We assume that there are $m$
boundary
faces and $s_i=0$ define them, where
$i=1,...,m$. Then $s_1 s_2\cdot \cdot \cdot s_m$
becomes a section of $\otimes^m_{i=1} L_i$.

Just as in the ${\bf P}^{n+1}$ case, we want to perturb
$s_1 s_2\cdot \cdot \cdot s_m$ by
a general section $W_0(z)$ of $\otimes^m_{i=1} L_i$ to obtain a smooth
hypersurface
(or with mild singularities):
$$W(z)=W_0(z)+\psi s_1 s_2\cdot\cdot \cdot s_{m}=0$$
and we will recover the union of the boundary hypersurfaces by
taking the $\psi \rightarrow \infty$ limit.  Note that since
we are considering a hypersurface inside
the toric variety as zeros of a section,
we can again give the toric
polytope an integral structure compatible with the choice of the
line bundle.
In order for this hypersurface to be Calabi-Yau
we need $\otimes^m_{i=1} L_i$ to be the same as the inverse of
the canonical line bundle $K^{-1}_{{\bf P}_{\Delta}}$ of
${\bf P}_{\Delta}$
in order to cancel the $c_1$. It
turns out that this condition can be rephrased in terms of the
integral polytope $\Delta$.

To do that we need to first introduce the idea of dual polytope. For any
integral polytope $\Delta $ containing the origin as an
interior point, its dual polytope $\nabla $ is
roughly the convex polytope bounded by hyperplanes $
a_{1}y_{1}+a_{2}y_{2}+...+a_{n}y_{n}=1$ for
$(a_{1},a_{2},...,a_{n}) $ any vertex of $\Delta .$ More precisely the
dual polytope is defined by
$$
\nabla =\left\{ ( y_{1},...,y_{n})
:x_{1}y_{1}+x_{2}y_{2}+...+x_{n}y_{n}\leq 1\quad  {\rm  for }\quad (
x_{1},...,x_{n}) \in \Delta \right\} .
$$
For example let $\Delta $ be the polytope with vertexes $( -1,-1)
,( -1,2) $ and $( 2,-1) $ in the plane corresponding
to ${\bf P}^2$ with a degree 3 bundle on it, then its dual
polytope $\nabla $ is bounded by $( 1,1) ,(- 1,0) $
and $( 0,-1) $ as shown below:

%

\bigskip

\centerline{\epsfxsize 2.truein \epsfysize 2.truein\epsfbox{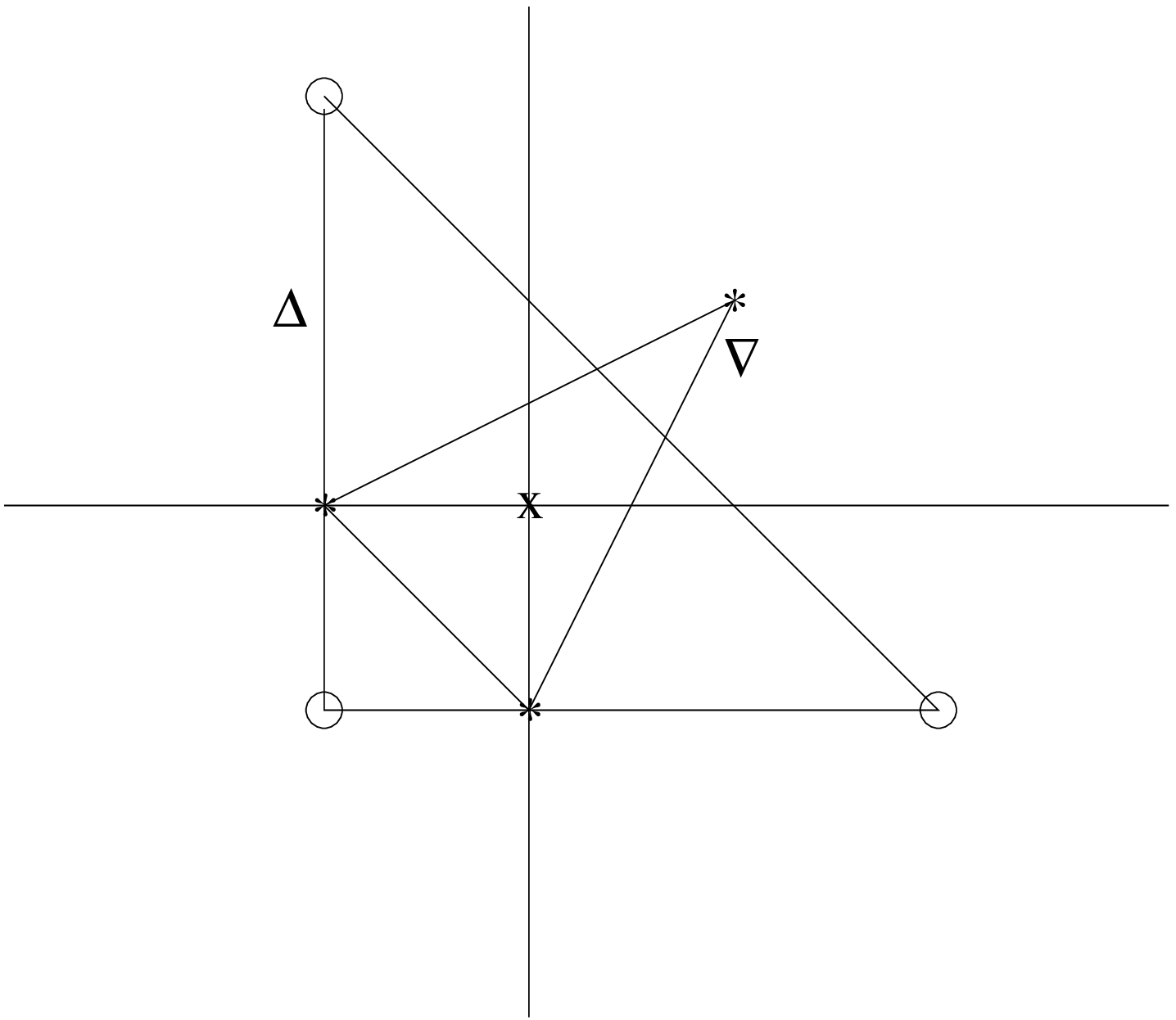}}
\leftskip 2pc
\rightskip 2pc
\noindent{\ninepoint\sl \baselineskip=8pt {\bf Fig.12}: {\rm Here
we are showing the integral reflexive polytope corresponding
to ${\bf P}^2$ (denoted by $\Delta$) and its dual denoted
by ${\nabla}$.  Note that for every face of $\Delta$ we get
a vertex of $\nabla$ and vice-versa.  The origin
connected to each vertex of $\nabla$ is orthogonal
to the corresponding dual face in $\Delta$.
Also note that
the origin is the only interior integral point.}}


In general the dual polytope of an integral polytope may not be integral
again, when this is the case, $\Delta $ will be called a reflexive polytope.
In this case, its dual polytope is also a reflexive polytope. For a
reflexive polytope, each vertex of $\Delta $ will correspond to a boundary
face of $\nabla $ and vice versa \baty
(this is illustrated in the figure for the ${\bf P}^2$ example above).
More generally for every $k$ dimensional face of the polytope
$\Delta$ there is a dual $n-k-1$ dimensional face of
$\nabla$.

In fact $\Delta $ being reflexive is equivalent to $K_{\bf{P}_{\Delta
}}^{-1}=\bigotimes_{i=1}^{m}L_{i}$ which guarantees the corresponding
hypersurface to be Calabi-Yau.  The deformation which we denoted by
$s_1...s_m$ will correspond to the origin for reflexive polytopes.
This can be easily seen to be the case for the ${\bf P}^2$ example
discussed before and turns out to be a general fact.

These constructions can be generalized to the case of varieties defined
by more equations.
Our discussions can be easily generalized
to these cases as well.  We will leave this to the reader.

\subsec{The Dual Toric Constructions}
If $\Delta$ is a reflexive polytope then ${\nabla }$ is also reflexive.
Therefore it defines another
toric variety ${\bf P}_{\nabla }$ and the zero of a general section $W_0'$
of  $K^{-1}_{{\bf P}_{\nabla }}$
will be Calabi-Yau.
This construction is proposed by Batyrev to
obtain the mirror of Calabi-Yau  hypersurface in ${\bf P}_{\Delta}$.
Again we can take the limit
as  $\psi' \rightarrow \infty$ of
$$W'(z)=W_0'(z)+\psi' t_1 t_2\cdot\cdot \cdot t_{m'}=0$$
to obtain the union of the boundary faces of ${\nabla }$ and each such face
corresponds to some $t_i=0$.

In the limit of  $\psi$ and $\psi' \rightarrow \infty$,
the geometries of the Calabi-Yau
hypersurfaces are described by the boundary of $\Delta$ and ${\nabla }$.
Each boundary
face of $\Delta$ will correspond to a vertex in ${\nabla }$ and vice versa.
This is a manifestation of
the $R \rightarrow 1/R$ duality of tori as we will see later in this
paper.

\subsec{Open Toric Varieties and their Dual}
As in our earlier discussions, we are also interested in non-compact
cases such as
$N( {\bf P}^2  )$,
as
they can form a local piece of a Calabi-Yau threefold.
Instead of just one space like $N( {\bf P}_\Delta  )$,
we can also have a union of several toric varieties intersecting each
other along toric subvarities. Examples of this kind include the
blown up of $A_n$ singularity where n ${\bf P}^1$ intersecting each
other in a linear manner as we already discussed.   A large
number of such examples has been considered in \kmv .

One can extend the
definition of duality for open toric varieties as well.
We first look at the dual polytope similar to the global case. Put
the origin inside the polytope $\Delta$ and consider all the rays
emanating from the origin and orthogonal
to each face.
If the boundary face is defined by $
a_1x_1+...+a_nx_n=c,$ then the corresponding ray will pass through the point $
( a_1,...,a_n) $.  If the polytope has $m$ faces then
 the collection of rays is characterized by $m$ integral
points $(a_1,...,a_n)$.  For example for the $A_n$ case discussed
before we get the collection of $(n+2)$ points in ${\bf Z}^2$ given by
$[(-1,1),(0,1),(1,1),...,(n,1)]$.   Similarly for $N({\bf P}^2)$ we get
the four points $[(1,1,1),(0,-1,1),(-1,0,1),(0,0,1)]$
 Note that the last entry in all these cases is
$1$.  This reflects the fact that if we fix the normal direction circle,
to shrink at a given point, as we go from one face to another, which
circle shrinks gets modified only by addition with a circle describing the
compact
pieces.   Thus the data of the dual object given by rays
will contain the same
information as a collection of points in a 1 dimensional lower
integral lattice consisting of $m$ integral points.
For the case of $A_n$ this will give us
$[(-1),(0),(1),...,(n)]$ which consists of $n+2$ integral points
along the line.  Similarly for the $N({\bf P}^2)$ case we get the points
$[(1,1),(0,-1),(-1,0),(0,0)]$.  Note that these are exactly
the same points defining the dual polytope for ${\bf P}^2$ together
with the interior point $(0,0)$.

\newsec{Branes and Toric Geometry}
There are some examples known where geometry can be replaced
with a configuration of branes.  These include M-theory
with $A_n$ singularity which correspond to $n$
D6 branes of type IIA \ref\wh{P.K. Townsend, Phys. Lett.
{\bf B350} (1995) 184.}\
type IIA/B over an $A_n$ singularity which
corresponds to type IIB/A with $n$ NS 5-branes of
type IIA/B \oogv , type
IIA/B  over a conifold, which is equivalent to 2 intersecting
NS 5 branes \ref\bsv{ M. Bershadsky, V. Sadov and C. Vafa, Nucl. Phys. {\bf
B463} (1996)
398.}  (see also
\ref\gib{J.P. Gauntlett, G.W. Gibbons, G. Papadopoulos and P.K. Townsend,
Nucl.Phys. {\bf B500} (1997) 133\semi
G. W. Gibbons, hep-th/9707232.}\ref\sena{A. Sen, hep-th/9707123.
}).  It is natural to try to
extend this dictionary to other singularities of geometry
(see for example
for one
such extension), and as it turns
out toric geometry is the right language for this purpose.
In fact those geometries which are locally
a toric space, can be realized via branes.  However
as we have mentioned before,
and will see below again, not all geometries have
toric realization.  In particular, in order to see some interesting
 geometries we have to consider hypersurfaces
(or complete intersections) in a higher dimensional toric
variety.  In such cases in general there is no known way
to associate a configuration of branes.  Thus it appears
that in geometry we have a more general approach in engineering
physical systems.

We have seen that toric geometries are essentially
trivial except for the fact that on some loci some
circles shrink.  The shrinking circles
are a source of charge of branes and so these loci
are naturally identified with branes of appropriate
type.  This is the basic link between toric geometry
and branes.

\subsec{M-theory on $S^1$ and $D6$ branes}

Let us consider the simple example of
the $A_1$ singularity $T^*{\bf P}^1$.  As noted above
M-theory on this space is equivalent to type IIA with
 $2$ units of $D6$ branes.  To see this note that the KK
monopoles of M-theory on a circle being equivalent to $D6$ branes
means that if we consider a supersymmetric geometry
in which the circle of M-theory shrinks to zero
size at some loci, we obtain the $D6$ branes.  Consider the toric geometry
above and identify the $S^1$ circle of $M$-theory with the $\theta$
action on $T^*{\bf P}^1$ discussed in example ii').  If we consider modding
out this space by the circle action
$$T^*{\bf P}^1/S_\theta$$
where the $S_\theta$ denotes the circle action associated with the
$\theta$ direction, we obtain a 3-dimensional geometry which
we can identify with the type IIA space.  Moreover the points on the
geometry where the $S_\theta$ has zero size, correspond to
$D6$ branes.  These are the points where the $a$ cycle vanishes
and from the toric diagram it is clear that this happens only
at two points on the ends of the interval representing
the ${\bf P}^1$. Similarly for M-theory on the $A_{n}$ geometry we find
it is equivalent to type IIA in the presence of $n+1$ $D6$ branes,
where the $n+1$ points correspond to the $n+1$ end points of the $n$
compact ${\bf P}^1$'s represented by the $n$-intervals in the toric diagram.

\subsec{M-theory on $T^2$ and $(p,q)$ 5-branes}

Above we have considered M-theory on $S^1$.  Let us now consider
M-theory on $T^2$. This theory is equivalent to type IIB on $S^1$.
 There are two cycles
on the $T^2$, and the Kaluza-Klein monopoles associated to the $(p,q)$
cycle of the $T^2$ corresponds to $(p,q)$ 5-branes of type IIB
(transverse to the compactified $S^1$ of type IIB).

This relation allows us to realize the local geometry
of Calabi-Yau threefolds which have toric realization
in terms of type IIB $(p,q)$ 5-branes.  Such geometries
will have a compact $T^2$ action corresponding to the 4-dimensional
compact local model, which we can mod out, and just as in the above
example realize them in terms of $(p,q)$ 5-branes.  For example
let us consider M-theory on
the Calabi-Yau threefold with a small ${\bf P^2}$.
Then the local model is as in example iii') above.
Considering modding out by the $T^2$ action corresponding
to the two finite circle actions on the ${\bf P}^2$.
Let $N({\bf P}^2)$ denote the ${\bf P}^2$ with the normal
bundle on top of it, and consider the 4-dimensional space
$$N({\bf P}^2)/T^2$$
where the $T^2$ denotes the lift of the action from the one
on ${\bf P}^2$.  This 4-dimensional space is trivial except for the
loci where a circle action of $T^2$ has fixed points--This happens
when the $a$ and $b$ cycle vanish and from the diagram in Fig. 11
we see that this occurs on the subspace shown below (with
the corresponding shrinking cycle $(a,b)$ indicated).
Note that when two faces in Fig. 11 meet, any combination of vanishing
cycles on either side vanishes on the intersection.


\bigskip

\centerline{\epsfxsize 2.truein \epsfysize 2.truein\epsfbox{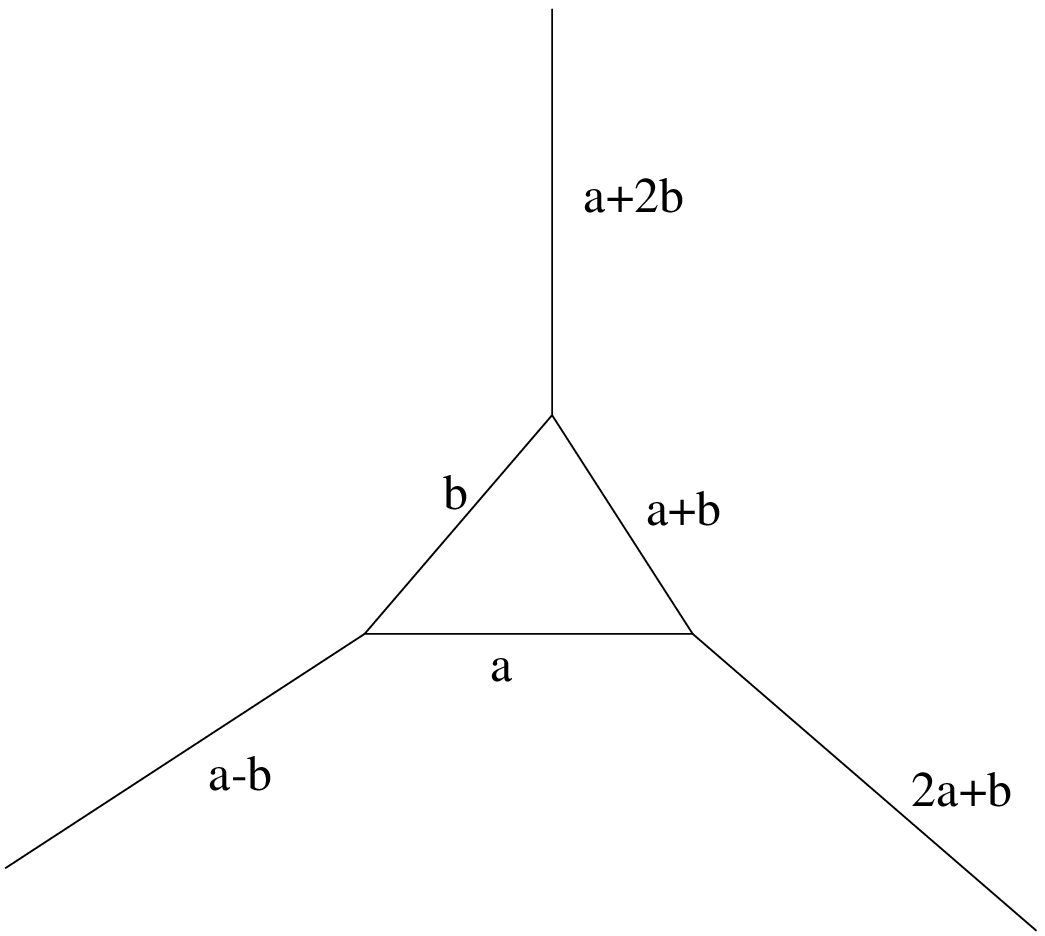}}
\leftskip 2pc
\rightskip 2pc
\noindent{\ninepoint\sl \baselineskip=8pt {\bf Fig.13}: {\rm The
brane realization of $N({\bf P}^2)$ in Calabi-Yau threefold.
All that has happened is that we have replaced the toric skeletons
with the corresponding $(p,q)$ 5-branes.}}


We thus interpret this as a type IIB geometry with the above
diagram giving the configuration of $(p,q)$ 5-branes
(where one extra $S^1$ has an arbitrary size on
the type IIB side).  This is
exactly the geometry proposed in \hanah\hanar\ for a critical
theory dual to the ${\bf P^2}$
shrinking in the Calabi-Yau.  Here we have explained this duality.
Moreover for every geometry proposed in \hanah\ we can
write down the geometric analog.

Note that the condition of $(p,q)$ charge conservation in \hanah\
gets mapped to the condition of $c_1=0$ for the local 3-fold.
At each vertex, using the $SL( 2,\bf{Z}) $ symmetry, we can
bring the local configuration to the standard basis $( 1,0) $ and
$( 0,1) $ in $\bf{Z}^{2}$ together with an external leg with
label $( p,q) .$  For $N( \bf{P}^{2}) $ to be
Calabi-Yau, as a line bundle over $\bf{P}^{2},$ it must be
$K_{\bf{P}^{2}}^{-1}$, i.e.,
the tensor product of the three line bundles corresponding to
the three edges of the standard 2 simplex. Now the origin is given by the
intersection of the two edges $z_{1}=0$ and $z_{2}=0.$ Therefore the $T^{2}$
action on the fiber over the origin is given by
$p\rightarrow \exp (-i\theta -i\phi ) p$ as in iii).
Hence the projection of this fiber to
the $( x,y) $ plane is the half line corresponding to the label
$( -1,-1) $ which is the same as the $( p,q) $ charge
conservation.

In fact for the models in \hanah\ the angles at which the
5-branes intersected was related to the $(p,q)$ charge.
Basically (at $\tau =i$) the $(p,q)$ 5-brane was
placed along the $(p,q)$ direction.  As we have said before
 this is very natural from toric geometry view point as well.
This is a beautiful interplay between branes and toric
geometry.

This dictionary we have found between geometry and branes will
explain some of the issues which were puzzling in
\hanah \hanar :  It was observed there that when one tries to construct
the brane version of the critical theory in 5 dimensions
corresponding to ${\bf P}^2$ blown up at more than 3 points
one gets certain puzzles.  One finds that the external lines
become parallel or interesecting (for more than
5 points blown up) in such cases.
This in particular
prevents their interpretation as a critical 5 dimensional theory.
This is puzzling
because ${\bf P}^2$ blown up at up to 8 points should lead
to critical theory \ref\mvii{D. R. Morrison
and C. Vafa, \nup 476 (1996) 437}\ref\wdel{E. Witten,
\nup 471 (1996) 195.}\ref\kdv{M. R. Douglas,
S. Katz and C.  Vafa, hep-th/9609071}\ref\mse{D. R.
Morrison and N. Seiberg, \nup 483
(1997) 229}.
Even though the case with parallel external
lines appear less harmful than the one with intersecting
external lines, some puzzles were raised even for this case
 in \hanar .  Note that this would limit the number of blow up
points to 3, if we were to realize it torically.

In order to resolve this puzzle
let us translate the condition of parallel external lines
to geometry:  If two parallel external lines bound an interval
in the toric polytope, then the $c_1$ evaluated on the ${\bf P}^1$
which is represented by the interval is zero.  To see this,
without loss of generality we can denote the 5-brane
charges on one end of the interval to consist
of $(-1,1)$ and $(0,1)$ internal line 5-branes meeting
the $(-1,0)$ external 5-brane, and the other end the
$(0,1)$, $(1,1)$ 5-brane meeting the $(-1,0)$ external
5-brane which is parallel to the other external line.
However the 2-dimensional complex
piece of this geometry is {\it exactly}
the same as the one we already studied, and the geometry
of ${\bf P}^1$ in ${\bf P}_\Delta$ is locally the same as $T^*{\bf P}^1$.
This implies that
$$\int_{\bf{P}^{1}}c_{1}(\bf{P}_{\Delta
}) =0$$
as is well known for the $T^*{\bf P}^1$ geometry.
Similarly when external parallel lines intersect, one
can show this implies that $c_1({\bf P}_\Delta)$ integrated
over the middle ${\bf P}^1$ is negative.

%

\bigskip

\centerline{\epsfxsize 2.truein \epsfysize 2.truein\epsfbox{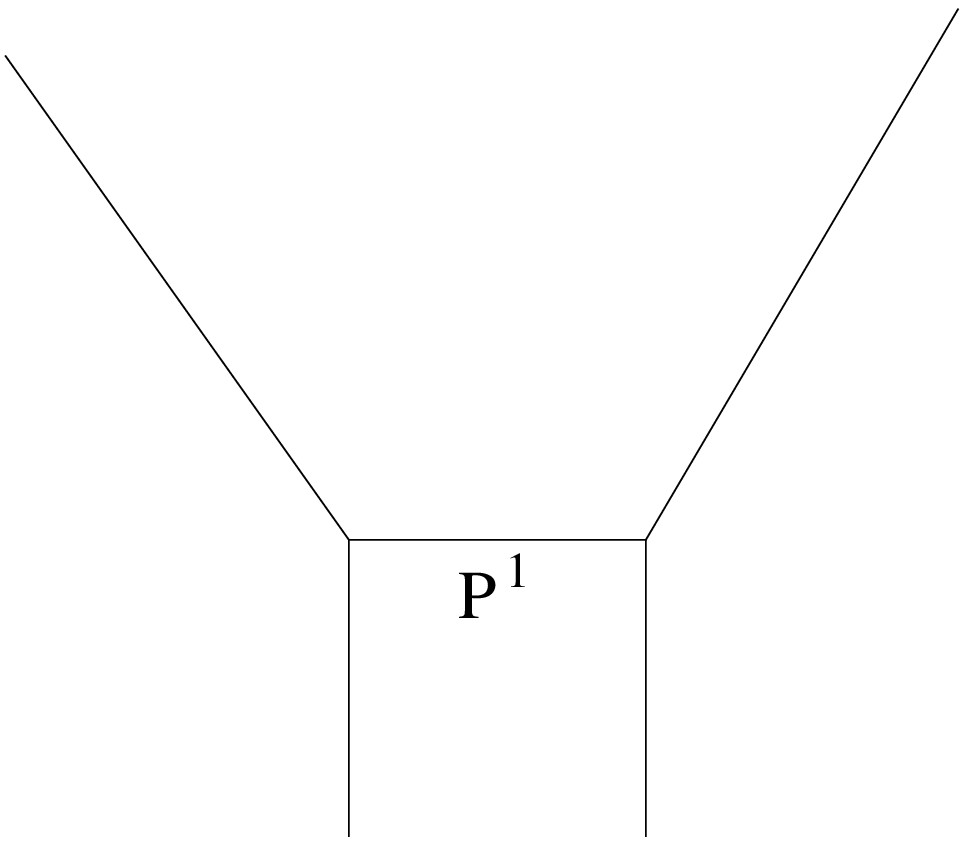}}
\leftskip 2pc
\rightskip 2pc
\noindent{\ninepoint\sl \baselineskip=8pt {\bf Fig.14}: {\rm When
we have parallel external lines the geometry in the neighborhood
of the middle ${\bf P}^1$
in the 2-complex dimensional base,
is essentially that of $T^*{\bf P}^1$.}}

However a submanifold can be shrunk within a background geometry if and
only if its normal bundle is negative by Grauert's criterion
\ref\mathref{H. Grauert, Uber Modifikationen und exzeptionelle
analytische Mengen, Math. Annalen, 146 (1962) 331.}. When the background
is a Calabi-Yau, that is  $c_1 =0$, then this is equivalent to
the submanifold having $c_1$ strictly positive\foot{
We would like to thank D. Morrison for a discussion on this point.}.
This means that even though
these geometries make sense locally they cannot be shrunk to zero
at finite distance in moduli.  In other words they do not
lead to conformal theories in 5 dimensions.

As mentioned before and reviewed above, the fact that with more than
3 points blown up ${\bf P}^2$ cannot be realized torically
is mathematically well known.   However as noted before
even between  3 and 8 points blown up ${\bf P}^2$ can shrink
in a Calabi-Yau, and can be realized in a higher dimensional
toric variety when we impose equations.
We are just learning that the brane realization
of quantum field theories
appear to be more limited than geometric engineering approach.
Or turning it around, we should try to understand what is the brane analog
of going to higher dimensions in geometry and imposing equations to decrease
the dimension back down.
There are some hints how this may be possible:
In particular there is a simple brane realization of ${\bf P}^2$ with
9 points blown up in terms of F-theory background, which involves
$(p,q)$ 7-branes compactified on a ${\bf P}^1\times S^1$
\mvii\wdel\ref\klemv{
A. Klemm, P. Mayr and C. Vafa, hep-th/9607139.}
(see \ref\bpsm{J. A. Minahan, D. Nemeschansky and N. P. Warner,
hep-th/9707149.}\ for an extensive study of BPS states
in this case).  However, one
should keep in mind that
the main question
is not whether a given geometry has a brane realization
or not, but more importantly whether it has a useful
brane realization.  In the above brane realization
one reverts to the geometric picture of M-theory on ${\bf P}^2$ with
up to 8 points blown up to extract physical results \klemv \bpsm\
(such as BPS spectrum).  Another (and perhaps more useful)
brane realization in this case may be
to consider a knotted configuration
of $(p,q)$ 5-branes and $(p,q)$ 7-branes piercing through them
(perhaps corresponding to blowing up points
{\it inside} the ${\bf P}^2$ triangle).  It should be interesting
to study such cases.

\subsec{M-theory on $T^3$ and (p,q,r) 4-branes}

There are many extensions of the above toric construction.
We will limit ourselves just to one more example, though
we believe our approach can be used in many different
contexts.

Consider M-theory on $T^3$.  Then we have the $SL(3)$ symmetry
as part of the $U$-duality group.  The KK monopoles will now be
labeled by a vector $(p,q,r)$ and will correspond
to a $4$-brane in the 7-dimensional geometry.  From the type IIB
perspective this corresponds to compactification on $T^2$ where
we consider a $(p,q)$ 5-brane wrapped around one of the circles
and bound to a KK monopole of charge $r$ around that circle.
There is also a type IIA description of the same object:  It corresponds
to compactification on $T^2$ with $r$ units of $D6$ brane
wrapped around $T^2$ and bound to KK monopoles of charge
$(p,q)$ on $T^2$.

Now just as a simple application, consider M-theory on a local
singularity of a Calabi-Yau 4-fold, yielding an
$N=2$ system in 3 dimensions.\foot{ These theories typically
have superpotentials generated. Here we will not worry
about whether there are such terms generated or not, and simply
consider the ``classical theory".  In fact for a shrinking
${\bf P}^3$ considered here there is a superpotential generated
\ref\wits{E. Witten, hep-th/9604030}.}.
 For example consider
${\bf P}^3$ shrinking inside a Calabi-Yau 4-fold.
(similar shrinking spaces can be considered and some
are equivalent to the models considered in \hanah )
Then the local model is the canonical bundle
over ${\bf P}^3$ which we denote by $N({\bf P}^3)$.
Then just as in the case of ${\bf P}^2$ considered above
we can realize the geometry of the toric polytope in terms
of the $(p,q,r)$ 4-brane intersecting
in a particular geometry in 3-dimensions.  Moreover
the intersection angles are exactly dictated by the
toric data, just as in the ${\bf P}^2$ case.
The figure of intersections look as follows.


\bigskip

\centerline{\epsfxsize 2.truein \epsfysize 2.truein\epsfbox{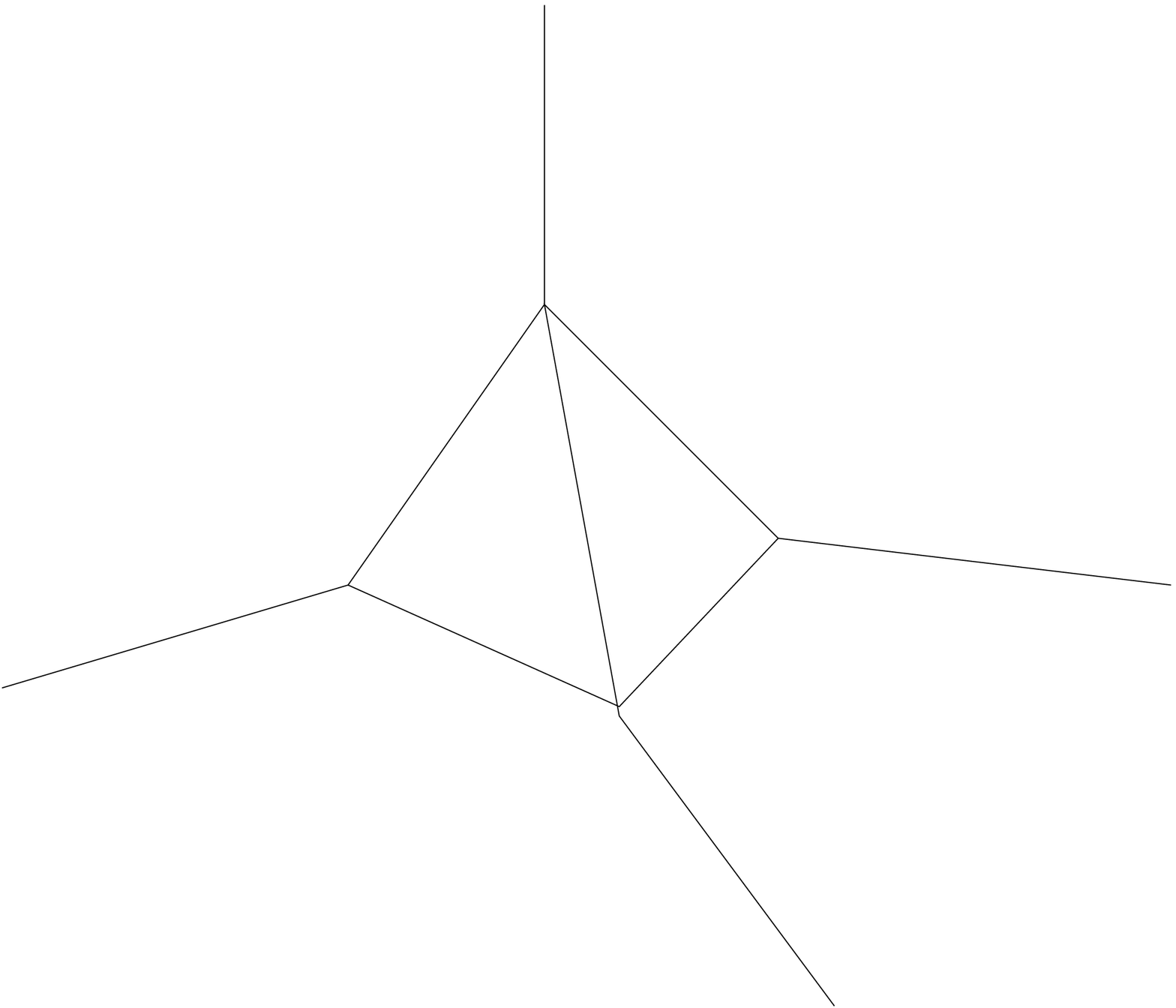}}
\leftskip 2pc
\rightskip 2pc
\noindent{\ninepoint\sl \baselineskip=8pt {\bf Fig.15}: {\rm The
brane realization of ${\bf P}^3$.  In addition to the 4 faces
of the tetrahedron corresponding to certain $(p,q,r)$ 4-branes,
there are 6 external
4-branes ending on each of the 6-edges of the tetrahedron.}}


Reduction of such cases to 2 dimensions and
local mirror symmetry in such cases has been recently considered
in \lerc .

\newsec{Intuitive explanation of mirror symmetry}
One of the most important
applications of toric geometry has been to mirror
symmetry.  In fact starting with the work of Batyrev \baty\
the mirror symmetry conjecture has been made quite systematic
using toric geometry.  Many attempts have gone into proving
mirror symmetry using this structure \ref\mpl{D. R. Morrison and M. R. Plesser,
 hep-th/9508107}\ref\szy{A. Strominger, S.-T. Yau and E. Zaslow, Nucl. Phys.
{\bf B479} (1996)
243.}\ref\morrison{D.R. Morrison, alg-geom/9608006.}.  The basic
idea being that the toric geometry means we have tori as fibers
and $R\rightarrow 1/R$ duality symmetry of each circle should
give rise to a simple description of mirror symmetry. Unfortunately
none of these approaches have been made complete.  An exception
to these cases involve
orbifolds for which mirror symmetry have
already been rigorously proven
to follow from $R\rightarrow 1/R$ symmetry \ref\vw{
C. Vafa and E. Witten, J. Geom. Phys. 15 (1995) 189.}.
Here we follow the same spirit of argument and find some
intuitive explanation of mirror symmetry again reducing
it to $R\rightarrow 1/R$.  Our approach will not be rigorous,
but we believe it makes mirror symmetry very plausible
and intuitive.  In particular we find a simple
explanation of Batyrev's construction for mirror
pairs.  We believe our approach can be generalized
to $(0,2)$ sigma models without much difficulty.

We divide our discussion into two parts.  One involving the global
case, where we consider compact Calabi-Yau manifolds realized
as hypersurfaces in toric geometry and the second one involving
what is called `local mirror symmetry' \kkv\kmv .  This latter one is the one
of
most interest in `solving' the $N=2$ gauge theories
in 4 dimensions.  The intuitive explanation that we will
find for this case makes the toric construction
as simply related to the geometry of the $N=2$ curves as
the one recently found in \brref\ but the geometry
appears to be distinct from it.
However, just as was the case for ${\bf P}^2$ with more
than 3 points blown up where we cannot just use a local toric model
to realize all such models.  In general we need to apply mirror symmetry in
situations as in the global case where
we need a higher dimensional space with some
equations imposed.  Again finding a brane analog for such cases
is challenging.
  Many such cases were addressed geometrically in \kkv\
and are a special limit of the global case considered below.

\subsec{global case}
Let us start with an example:  Consider ${\bf P}^{n+1}$ with
degree $n+2$ polynomial.  As discussed before this gives
rise to a Calabi-Yau $n$-fold.  At a particular singular
complex structure
limit this Calabi-Yau becomes toric and is identified
with the boundary of the $n+1$ dimensional polytope.
There are $n+2$ faces, over each of which we have
a $T^n$ fiber (one of the circles have already shrunk
by restricting to the boundary).  Now, if we perform T-duality
for each circle of $T^n$ replacing it by $1/R$, we should get a mirror
Calabi-Yau.  However,  note that where $k$ faces
meet
the $T^n$ shrinks to $T^{n-k}$ and which $T^{n-k}$ we get
depends on which faces meet.  Applying T-duality
on such loci leads to the following picture:
not only it restores the shrunk circles to big circles,
but makes them more dominant than the finite size circles.
In order to get a better picture, let us assume the zero size
circles have finite size $\epsilon$ and at the end we let $
\epsilon \rightarrow 0$.  Moreover let us change the Kahler
class of the Calabi-Yau so that all the lengths are rescaled
by the large factor $1/\epsilon$.  Now the cycles which were finite
size become of order of $1/\epsilon$ and the ones which were shrinking
become finite size.  Now we apply the T-duality.  In this case
all the finite size circles shrink and all the ones which were
previously finite size become of size one.
One gets a new space which in fact by definition is
the boundary of the dual polytope!  Note that {\it the construction
of the dual polytope where there is a Poincare duality on the boundary
of polytope is now interpreted very physically as the exchange
between the regions where the circles were small with
regions where the circles are big}.

Note that in the original manifold we have taken
{\it both} the Kahler and complex structures large.
Mirror transform will tell us that we should land on the
mirror manifold which is again corresponding to the large complex
class and Kahler class.  This is consistent with the fact that in order
to get the boundary of the dual polytope as the mirror, we are
at the large complex structure of the mirror which is mirror to the
large Kahler class of the original manifold. It is no accident
that we had to take the large Kahler structure of the original
manifold to get a simple description of the mirror as a degenerate
toric manifold.

At the large Kahler and complex structure limit, we can also identify, up to
first order, the Kahler moduli of the original manifold with the complex
moduli of its mirror and vice versa in more
detail.
 This would give a realization of
the monomial-divisor mirror map as proposed in \ref\mmir{P.S.
Aspinwall, B.R. Greene and D.R. Morrison, Internat. Math. Res. Notices
(1993) 319.}.
The arguments we give are parallel to those recently given in
\kolet \yan \hanar\ in connection with the M-theory realization
of $N=2$ gauge systems in 4 dimensions.

Let us illustrate how this works:
Suppose that $X$ is the Calabi-Yau hypersurface defined by $W\left( z\right)
=0$ inside a toric variety $\bf{P}_{\Delta }$ with $\Delta $ a reflexive
polytope as in section 2.2. We consider Kahler metric on $X$ which comes
from restriction of Kahler metric on $\bf{P}_{\Delta }.$ Each boundary
face of $\Delta $ will determine a line bundle
whose curvature is a closed two form on $\bf{P}_{\Delta }.$ Suitable
positive combination of them will give us general toric
Kahler metric on $\bf{P}_{\Delta }.$ Explicit description of these
Kahler forms are given in
\ref\torkah{V. Guillemin, Kahler structures on toric varieties
J. Diff. Geom. 40 (1994) 285.}.
In general cohomology classes of these Kahler forms could be dependent.
For example in the case of $K3$ surface in $\bf{P}^{3}$, there is only one
such Kahler form up to scaling factor.

In the $R\rightarrow 1/R$ duality, before we scale the Kahler class by the
global factor $1/\epsilon ,$ if we take the Kahler metric to be one which
is dominated by one of the boundary faces of $\Delta $, then in the mirror,
the corresponding face will become a vertex of $\nabla .$
As noted before
 any integral
point in the mirror reflexive polytope $\nabla$
 will correspond to a monomial
which is in fact a section $s$
of  $K_{\bf{P}_{\nabla }}^{-1}$ and which
can be used to deform the complex structure of the mirror.
Let $y_i$ with $i=1,...,n$ denote the coordinates of the mirror
toric variety.  Then as noted  the deformation corresponding to the origin
of ${\nabla}$ corresponds to the monomial $1$.  On the other
hand let $(a_1,...a_n)$ denote a vertex on $\nabla$.  To that we can
associate the monomial deformation on the
mirror of the form $y_1^{a_1}...y_n^{a_n}$.
Now let us consider the limit in complex deformation
of the mirror  in which these two terms have
large coefficients.  Then the equation defining the
manifold in the mirror is effectively
dominated by
$$\alpha + \beta
y_1^{a_1}...y_n^{a_n}=0$$
for some $\alpha, \beta$.
Notice that some of $a_i$'s can be negative and setting $\alpha$
goes to infinity gives the singular Calabi-Yau corresponding to
the union of boundary faces of $\nabla$. Now when we deform the
complex structure away from this singular point using the monomial
deformation by $y_1^{a_1}...y_n^{a_n}$.
then the manifold becomes dominated in this
limit by
\eqn\manif{y_1^{a_1}...y_n^{a_n}={\rm const.}}
Given that in our derivation of mirror symmetry
this vertex was associated to a face of the original polytope
$\Delta$ we would expect that this large complex structure
limit should be mirror to large Kahler class limit for this
particular divisor. But in this limit the manifold is dominated
by that face, whose real section corresponds to
\eqn\mie{\sum_{i} a_i x_i={\rm const.}}
As far as mirror symmetry is concerned the base of the geometry
for the manifold and the mirror are identified, thus $x_i$
should be related to real part of $y_i$.  In fact comparing
\manif\ with \mie\ we see that the natural identification of
$$y_i=e^{x_i}$$
will make \mie\ and \manif\ identical.  We have thus found
a simple physical explanation of the divisor/monomial mirror map.
In other words the complex deformation mirror to the Kahler
deformation controlling the size of a divisor is the coefficient
of the monomial associated to the vertex mirror to the divisor.

\subsec{local case}

Now consider a local toric model of Calabi-Yau, such
as $N({\bf P}^n)$. We wish to find its mirror, i.e.
a geometry whose complex structure is mirror to the
Kahler class of $N({\bf P}^n)$.  We can try to repeat an argument
similar to the above as in the global case, but we will try
to take a short cut.  This arises because some of the
data in the original manifold is not necessary in defining it.
In fact this was already reflected in our discussion of the dual
object associated in these cases.

The data characterizing the Kahler geometry of the model
in these non-compact cases is concentrated on the subspace
of polytope of dimension $n-1$.  For example for $A_k$ this will be
$k+2$ points which is also related to the fact that
we found
that the natural dual object in this case will consist of $k+2$
successive points on an integral
lattice in one dimension. In the case of the $N({\bf P}^2)$
this is a one dimensional graph together with three external lines, which
constitute the $(p,q)$ 5-branes, together with the $S^1$ fibers
on top.  In other words the data of the local geometry can be
reconstructed by a 1-dimensional singular object consisting
of $3 {\bf P}^1$'s meeting along a triangle and with $3$
half ${\bf P}^1$'s coming out from the vertices:


\bigskip

\centerline{\epsfxsize 2.truein \epsfysize 2.truein\epsfbox{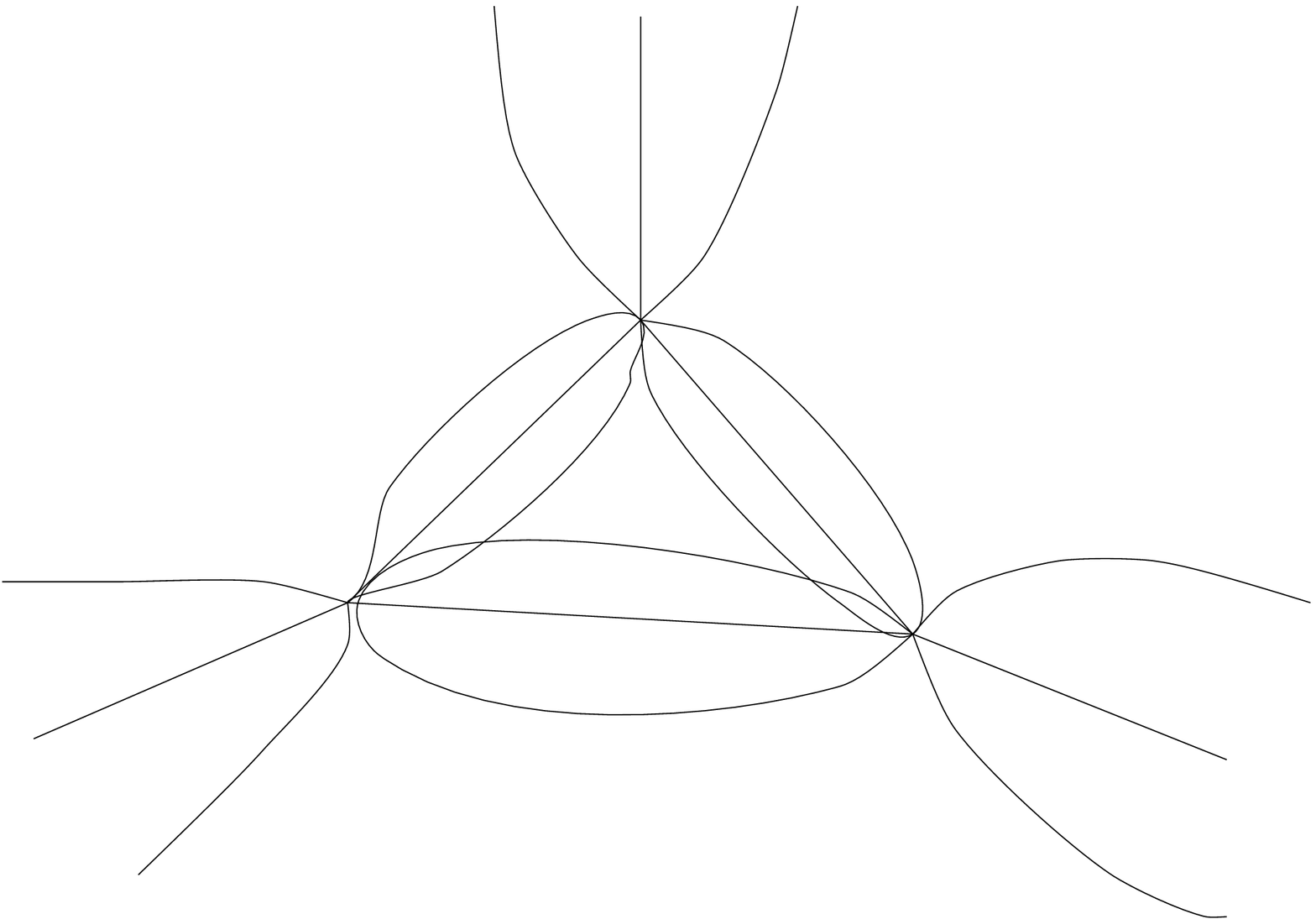}}
\leftskip 2pc
\rightskip 2pc
\noindent{\ninepoint\sl \baselineskip=8pt {\bf Fig.16}: {\rm
The toric realization of $N({\bf P}^2)$ is captured by
three spheres intersecting each other as well as three
external half-spheres as shown.  The elliptic curve
is visible in the picture as the fattened triangle in the middle.}}


More generally this gives the $n-1$ complex dimensional object
which is a $T^{n-1}$ toric fiber over an $n-1$ dimensional collection
of faces.
We now apply mirror symmetry to this situation by acting
on the $T^{n-1}$ fibers with $R\rightarrow 1/R$.   We can
trust this action away from the singular points where the curves meet.
Note that in the case of $N({\bf P}^2)$ we get a Riemann surface
and the endpoint of the external lines get mapped to a point on the surface
(because the circle infinitely big is mapped to a tiny circle).  So we obtain
an elliptic curve with 3 punctures.

To write the local
complex geometry of the mirror we can use quite generally
exactly the same idea as in the global case.
Namely consider the set of integral points which will be $m$ points
in a lattice of dimension $n$.  Recall that these are in one to
one correspondence with rays orthogonal to faces.
 Consider the complex space
$$\sum_{r=(a_1,...a_{n})\in {\rm space\ of\ rays }}
c_ry_1^{a_1}...y_{n}^{a_{n}}=0$$
which gives a space of complex dimension $n-1$ as expected above.
To make this application of mirror symmetry more intuitive
again we can extend the analog of $R\rightarrow 1/R$ duality
face by face.  Namely recall that the original polytope
is characterized by the projections of its faces onto an $n$
real dimensional space which consists of $n-1$
dimensional skeletons.   Consider one such face of the skeleton.
This will be bounded by two $n$ dimensional faces of the original
polytope which are associated with two monomials
$y_1^{a_1}...y_{n}^{a_{n}}$ and $y_1^{b_1}...y_{n}^{b_{n}}$.
 Now if we consider the limit of Kahler classes
where a particular  $n-1$ dimensional skeleton is large.  This
will be mirror to the complex deformation where the two monomials
have large complex coefficients.  In this limit the equation
for the mirror gets dominated by
$$\alpha y_1^{a_1}...y_{n}^{a_{n}}+
\beta y_1^{b_1}...y_{n}^{b_{n}} =0$$
which gives
$$y_1^{a_1-b_1}...y_{n}^{a_{n}-b_{n}}={\rm const.}$$
Again as in the global case since the $R\rightarrow 1/R$ does not
act on the real part this should also give the real part of the space
which is the skeleton.  If
we identify $y_i={\rm exp}(x_i)$ we obtain
the equation of the skeleton
$$\sum_{i} (a_i-b_i)x_i= {\rm const.}$$
as expected.
Just as an example if we consider the $N({\bf P}^2)$ case we get
(using the definition of the rays discussed in section 2)
$$ a +by_1+c y_2+{d\over y_1y_2}=0$$
as the local model for the mirror manifold.
Another example involves fibering $A_n$ spaces over ${\bf P}^1$
(the generalization of which to many examples
was already considered in detail in \kmv\
and the complex moduli including the coupling
can be viewed as the radius dependence of the 5 dimensional
critical theories compactified on a circle \ref\lwn{
A. Lawrence and N. Nekrasov, hep-th/9706025.}).
In this case we get the figure
%

\bigskip

\centerline{\epsfxsize 2.truein \epsfysize 2.truein\epsfbox{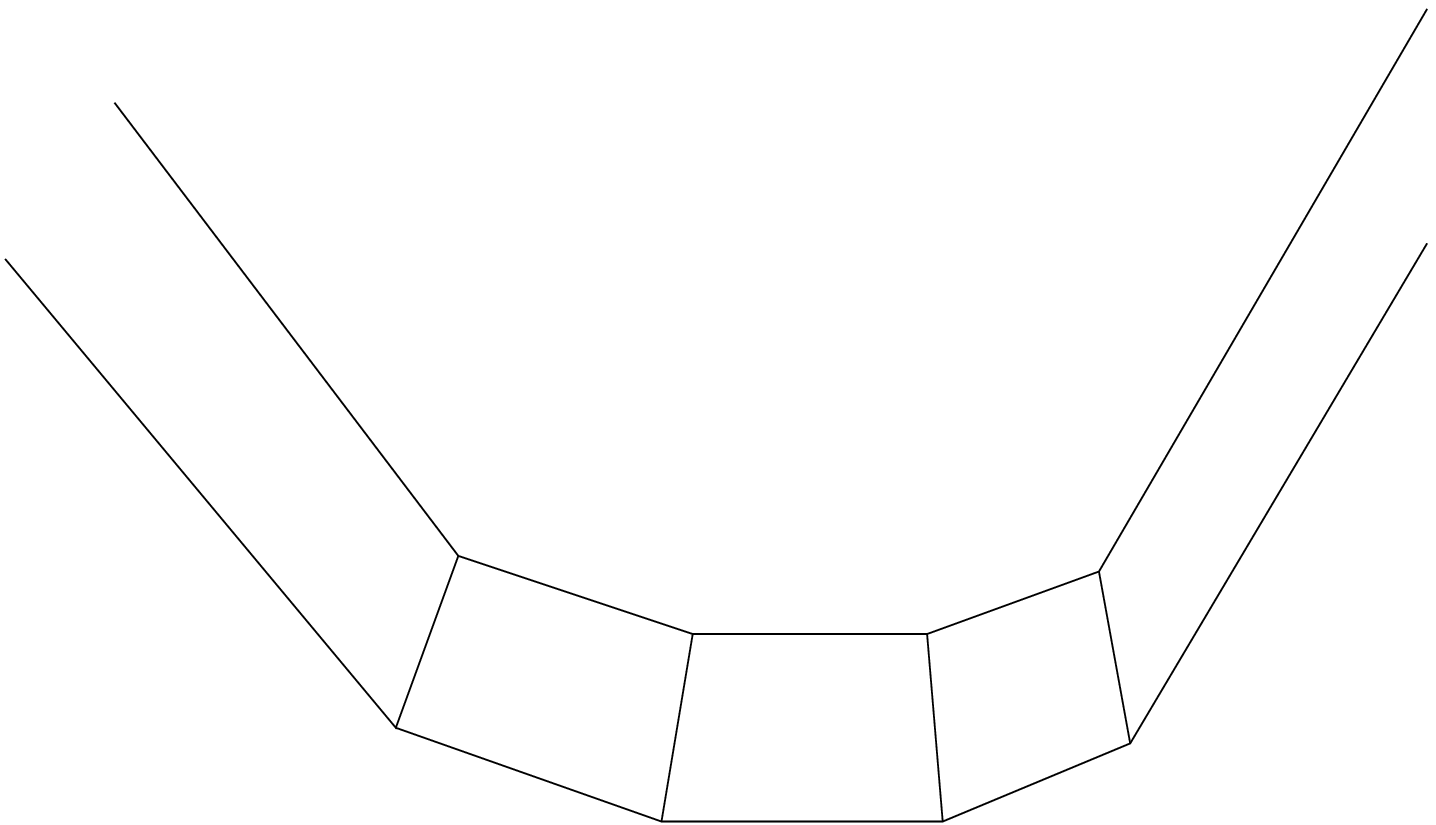}}
\leftskip 2pc
\rightskip 2pc
\noindent{\ninepoint\sl \baselineskip=8pt {\bf Fig.17}: {\rm The toric
realization of $SU(n)$ gauge theory produces a `visible'
genus $n-1$ Riemann surface as its skeleton.  The short
direction of the `ladder' in the above figure corresponds
to a ${\bf P}^1$ and the long direction corresponds to the blown up of
an $A_{n-1}$ space.}}


Here we see the $N=2$ Riemann surface very visibly
as the skeleton of the toric
graph.  Similar observations have been recently made
in connection with the M-theory approach \kolet \yan \hanar .

We would like to thank S. Katz, P. Mayr and S.-T. Yau for valuable
discussions.

The research of C.V. was supported in part by NSF grant PHY-92-18167.

\listrefs

\end